\documentclass[aps,prd,superscriptaddress,preprintnumbers,nofootinbib,notitlepage]{revtex4-2}
\usepackage[latin9]{inputenc}
\usepackage{bm}
\usepackage{amsmath}
\usepackage{amsthm}
\usepackage{amssymb}
\usepackage[colorlinks=true]{hyperref}

\newcommand{\dd}{\mathrm{d}}
\newcommand{\Mpl}{M_\mathrm{Pl}}

\begin{document}

\preprint{YITP-20-157, IPMU20-0126}

\title{Minimal Theory of Bigravity: construction and cosmology}

\author{Antonio De Felice}
\email{antonio.defelice@yukawa.kyoto-u.ac.jp}
\affiliation{Center for Gravitational Physics, Yukawa Institute for Theoretical Physics, Kyoto University, 606-8502, Kyoto, Japan}
\author{Fran\c{c}ois Larrouturou}
\email{francois.larrouturou@iap.fr}
\affiliation{Institut d'Astrophysique de Paris, UMR 7095, CNRS, Sorbonne Universit{\'e},\\ 98\textsuperscript{bis} boulevard Arago, 75014 Paris, France}
\author{Shinji Mukohyama}
\email{shinji.mukohyama@yukawa.kyoto-u.ac.jp}
\affiliation{Center for Gravitational Physics, Yukawa Institute for Theoretical Physics, Kyoto University, 606-8502, Kyoto, Japan}
\affiliation{Kavli Institute for the Physics and Mathematics of the Universe (WPI), The University of Tokyo Institutes for Advanced Study, The University of Tokyo, Kashiwa, Chiba 277-8583, Japan}
\author{Michele Oliosi}
\email{michele.oliosi@yukawa.kyoto-u.ac.jp}
\affiliation{Center for Gravitational Physics, Yukawa Institute for Theoretical Physics, Kyoto University, 606-8502, Kyoto, Japan}

\date{\today}

\begin{abstract}
Following the path of minimalism in alternative theories of gravity, we construct the ``Minimal Theory of Bigravity'' (MTBG), a theory of two interacting spin-2 fields that propagates only four local degrees of freedom instead of the usual seven ones and that allows for the same homogeneous and isotropic cosmological solutions as in Hassan-Rosen bigravity (HRBG). Starting from a precursor theory that propagates six local degrees of freedom, we carefully choose additional constraints to eliminate two of them to construct the theory. Investigating the cosmology of MTBG, we find that it accommodates two different branches of homogeneous and isotropic background solutions, equivalent on-shell to the two branches that are present in HRBG. Those branches in MTBG differ however from the HRBG ones at the perturbative level, are both perfectly healthy and do not exhibit strong coupling issues nor ghost instabilities. In the so-called self-accelerating branch, characterized by the presence of an effective cosmological constant, the scalar and vector sectors are the same as in General Relativity (GR). In the so-called normal branch, the scalar sector exhibits non-trivial  phenomenology, while its vector sector remains the same as in GR. In both branches, the tensor sector exhibits the usual HRBG features: an effective mass term and oscillations of the gravitons. Therefore MTBG provides a stable nonlinear completion of the cosmology in HRBG.
\end{abstract}

\maketitle

\section{Introduction and motivations} 
\label{sec:intro}

Among the clouds that obscure modern cosmology is the observation of an accelerated expansion of our Universe~\cite{DE_obs_1,DE_obs_2}.
If such mechanism can be phenomenologically explained by introducing an \emph{ad hoc} cosmological constant, it can also be addressed by a more fundamental infra-red modification of General Relativity (GR), for example by adding a small mass to the graviton.
The required mass to account for such accelerated expansion is extremely light (of order $10^{-33}$ eV), still far from the current upper bounds imposed by the detection of gravitational waves from a coalescing binary of neutron stars~\cite{GW170817_tests}.
Independently of this problem, and from a purely theoretical point of view, the idea that the graviton could bear a mass was first introduced in the seminal work of Fierz and Pauli~\cite{Fierz_Pauli}, where they constructed a viable massive gravity theory, at the linear level.
The healthy (i.e.\ ghost-free) completion at the fully non-linear level of Fierz and Pauli's idea has only been reached more than seventy years later by de Rahm, Gabadadze and Tolley~\cite{dRGT_1,dRGT_2}, and thus is dubbed ``dRGT massive gravity.'' As expected for a theory describing a massive spin-2 particle, it propagates five degrees of freedom (\emph{dof}). Crucial to this theory is the presence of a frozen, fiducial metric that combines with the dynamical one to give the graviton mass term.

Despite this great theoretical achievement, dRGT massive gravity suffers from the lack of stable purely Friedmann-Lema\^{i}tre-Robertson-Walker (FLRW) cosmologies~\cite{dRGT_no_FLRW}, which opened the door to many extensions.
A rather natural way of extending such theory is the so-called Hassan-Rosen bigravity (HRBG) where the fiducial metric acquires a canonical Einstein-Hilbert kinetic term, and thus is promoted to be dynamical~\citep{Hassan_Rosen}. This theory can be seen as describing the interaction of a massive graviton and a massless one, and thus propagates seven \emph{dof}. But this conceptually simple extension suffers from a gradient-type instability~\cite{bigrav_inst_1,bigrav_inst_2}. This instability can nevertheless be cured by carefully choosing the hierarchy of Planck scales~\cite{bigrav_Planck_hierarchy} or by adding a chameleon-like potential~\cite{Cham_bigrav_OP,Cham_bigrav_cosmo}.
Apart from bigravity, the search for ``beyond-dRGT'' models is still very active, and we can cite for example the ``Generalized Massive Gravity'' framework~\cite{GMG_OP,GMG_cosmo_1,GMG_cosmo_2,Non_minimal_coupling}, where the mass parameters are promoted to functions of the Stueckelberg fields. 

Nevertheless, all those theories contain five or more \emph{dof}, including non-tensorial ones, while GR only contains two tensor \emph{dof}. 
From a phenomenological point of view, the forty-five-year-long observation of the binary pulsar PSR B1913+16 has put stringent bounds on the energy radiated in scalar polarization, to be less than 1\% of the total energy radiated away~\cite{Will_pulsars}. 
In addition, the direct detection of gravitational radiation favors tensor-only polarization~\cite{GW170814_pol}, although the current configuration of the ground-based detectors is not very efficient for the purpose of discriminating polarizations and we will have to wait for space-borne ones to have better constraints~\cite{pol_LISA}.
From a theoretical point of view, it is also more credible to build theories that avoid unnecessary \emph{dof}.
This minimalistic approach has been applied to massive gravity, and a ``Minimal Theory of Massive Gravity'' (MTMG), propagating only 2 tensor \emph{dof} was constructed in~\citep{MTMG_OP}.
Such minimalism naturally has a cost: Lorentz invariance has to be explicitly broken down to the group of spatial rotations, albeit only weakly, by the mass term.
This model contains two branches of FLRW solutions, including a self-accelerating one~\cite{MTMG_pheno}, healthy static stars and black holes~\cite{MTMG_BH} and achieves better fit with the redshift distorsion data and the integrated Sachs-Wolfe-galaxy cross-correlated data than the current standard model~\cite{MTMG_RSD,MTMG_ISW}. More recently, nonlinear dynamics of MTMG was also studied by means of N-body simulations~\cite{MTMG_Nbody}. 
Nevertheless, while its FLRW solutions are free from the problems of dRGTs ones, they require a temporal dependence (or the dependence on the temporal component of Stueckelberg fields in the covariant formulation) of the fiducial metric as a part of the definition of the theory. 
To avoid such dependence, an extension with a single quasi-dilatonic scalar field was constructed~\cite{MQD_OP,MQD_Horndeski} and shown to have perfectly stable cosmologies~\cite{MQD_pheno}.
Note that this principle of minimalism has also been applied to generic constructions of alternative theories of gravity, escaping the Lovelock's theorem by the breaking of Lorentz invariance at cosmological scales, and has shown very interesting phenomenological implications~\cite{MMTG_OP,MMTG_Carballo_Rubio,MMTG_Lin,MMTG_Hamiltonian,MMTG_pheno,MMTG_Planck,MMG2_BH,Yao:2020tur,MMG2_weakening}.

If the addition of a quasi-dilatonic scalar field succeeded in implementing viable cosmologies, it seems more natural to promote the fiducial metric to be dynamical, thus constructing a ``Minimal Theory of Bigravity,'' which is the aim of the present work. 
We thus seek for a theory containing two interacting spin-2 fields, which propagates only four degrees of freedom. 
Inspired by the construction of MTMG, we start our procedure with a precursor, diffeomorphism-breaking theory that propagates six \emph{dof} (of which two are scalars).
Adding carefully chosen, extra constraints, we then reduce the number of \emph{dof} to be at most four.
By investigating its perturbations on a cosmological background, we then show that the model contains at least four gravitational \emph{dof}, thus concluding that we indeed have the required number of \emph{dof}. 

This work is organized as follows: in Sec.~\ref{sec:precursor} we present the precursor theory and count the number of its propagating degrees of freedom. From this precursor theory, we construct MTBG in Sec.~\ref{sec:MTBG} and discuss the number of \emph{dof} it contains. Its cosmology is investigated in Sec.~\ref{sec:cosmo}, and we conclude in Sec.~\ref{sec:concl}.
App.~\ref{app:Ham_pre} and~\ref{app:Ham_MTBG} present the details of the Hamiltonian analysis of the precursor theory and MTBG respectively, and App.~\ref{app:Vielbein} gives the expression of the theory in the vielbein language.
App.~\ref{app:dic} provides a dictionary between notations in the present paper and those in \cite{bigrav_viable_cosmo}. 
Finally, App.~\ref{app:hamil_back} discusses the FLRW background equations of motion in the Hamiltonian formulation.

\section{Precursor theory}
\label{sec:precursor}

\subsection{Precursor action}

The first step towards the construction of the MTBG is the definition of a precursor action, containing two dynamical metrics, dubbed $g_{\mu\nu}$ and $f_{\mu\nu}$, that interact through a dRGT-inspired, 4D-diffeo-breaking mass term. In order to construct such interaction, let us introduce the ADM decompositions of the two metrics
\begin{equation}
\begin{aligned}
& \dd s_g^2 =
g_{\mu\nu}\dd x^\mu \dd x^\nu =
 -N^2\,\dd t^2+\gamma_{ij}\left(\dd x^i+N^i\,\dd t\right)\left(\dd x^j+N^j\,\dd t\right)\,,\\
& \dd s_f^2 =
f_{\mu\nu}\dd x^\mu \dd x^\nu =
 -M^2\,\dd t^2+\phi_{ij}\left(\dd x^i+M^i\,\dd t\right)\left(\dd x^j+M^j\,\dd t\right)\,.
\end{aligned}
\end{equation}
An important point is that we will hereafter work in the so-called ``unitary gauge,'' i.e.\ without explicit covariantization \emph{via} the introduction of Stueckelberg fields. While such covariantization is straightforward, it suffice to work in the unitary gauge for the purpose of the present paper, i.e.\ construction of the theory and study of cosmology. 

In dRGT massive gravity and HRBG, the interaction term is constructed as a linear combination of the elementary symmetric polynomials of the ``square-root matrix'' $\mathbb{K}^\mu{}_{\nu}$, such that $\mathbb{K}^\mu{}_{\rho}\mathbb{K}^\rho{}_{\nu} = g^{\mu\sigma}f_{\sigma\nu}$. As we want an interaction term that breaks one full copy of space-time diffeomorphism as well as the temporal component of the second copy, we will build it from a similar ``square-root matrix,'' but constructed instead out of spatial metrics, namely $\mathfrak{K}^p_{\ q}$ (and its inverse $\mathcal{K}^p_{\ q}$) that obeys
\begin{equation}\label{eq:gothic_K_def}
\mathfrak{K}^p_{\ s}\mathfrak{K}^s_{\ q} = \gamma^{pr}\phi_{rq}\,,
\qquad \text{and} \qquad
\mathcal{K}^{\ s}_q\mathcal{K}^{\ p}_s = \gamma_{qr}\phi^{rp}\,.
\end{equation}
where $\gamma^{ij}$ and $\phi^{ij}$ are the inverses of $\gamma_{ij}$ and $\phi_{ij}$, respectively. 
From those matrices, we define the 3-dimensional elementary symmetric polynomials, (with the brackets denoting the trace),
\begin{equation}
\begin{aligned}
& e_0\left(\mathfrak{K}\right) = 1\,,\\
& e_1\left(\mathfrak{K}\right) = \left[\mathfrak{K}\right]\,,\\
& e_2\left(\mathfrak{K}\right) = \frac{1}{2}\left(\left[\mathfrak{K}\right]^2-\left[\mathfrak{K}^2\right]\right)\,,\\
& e_3\left(\mathfrak{K}\right) = \det\left(\mathfrak{K}\right)\,,
\end{aligned}
\end{equation}
and similarly for $e_n\left(\mathcal{K}\right)$.
From those, we define the action of the precursor theory
\begin{equation}\label{eq:precursor_action}
\mathcal{S}_\text{pre} =
\frac{\Mpl^2}{2}\int\!\dd^4x \left\lbrace
\sqrt{-g} \mathcal{R}\left[g\right]
+ \alpha^2 \sqrt{-f} \mathcal{R}\left[f\right]
- m^2\sum_{n=0}^3\left[\sqrt{-g}c_{4-n}e_n\left(\mathfrak{K}\right)+\sqrt{-f}c_{n}e_n\left(\mathcal{K}\right)\right]\right\rbrace\,,
\end{equation}
where $\mathcal{R}\left[g\right]$ is the four-dimensional Ricci scalar constructed out of $g_{\mu\nu}$, $\alpha$ is the ratio between the Planck masses of the two sectors, $m$ and the set $\{c_n\}_{n=0..4}$ are coupling constants. Naturally there is a redundancy as one of the $c_n$ can be absorbed in $m^2$, but we will keep it as it is for simplicity.
This precursor action can be obtained from the usual HRBG in vielbein formulation, when imposing the ADM form for the vielbeins.
Note that similarly, the precursor action of MTMG was obtained from the usual dRGT massive gravity restricted to ADM-vielbeins~\cite{MTMG_OP}.

\subsection{Dirac procedure of the precursor theory}

Let us now count the number of \emph{dof} present in the precursor theory~\eqref{eq:precursor_action}, through a Dirac procedure~\cite{Dirac}.

As we are going to see, the lapses and shifts do not enter the mass term and, consequently, they enter only linearly in the Hamiltonian of the system. As a result, they can be simply treated as Lagrange multipliers. We thus start with the twenty-four dimensional phase space spanned by the spatial metrics and their conjugate momenta, defined as usual:
\begin{equation}
\pi^{ij} \equiv \frac{\delta \mathcal{S}_\text{pre}}{\delta \partial_t\gamma_{ij}} = \frac{\Mpl^2}{2}\sqrt\gamma\left(K^{ij}-K\gamma^{ij}\right),
\quad \text{and} \quad
\sigma^{ij} \equiv \frac{\delta \mathcal{S}_\text{pre}}{\delta \partial_t\phi_{ij}} = \frac{\alpha^2\Mpl^2}{2}\sqrt\phi\left(\Phi^{ij}-\Phi\phi^{ij}\right),
\end{equation}
where we have introduced the two extrinsic curvatures
\begin{equation}
K_{ij} = \frac{1}{2N}\left(\partial_t\gamma_{ij} -2\mathcal{D}_{(i}N_{j)}\right),
\qquad \text{and} \qquad
\Phi_{ij} = \frac{1}{2M}\left(\partial_t\phi_{ij} -2\mathfrak{D}_{(i}M_{j)}\right),
\end{equation}
$\mathcal{D}_i$ is the covariant derivative compatible with $\gamma_{ij}$, whereas $\mathfrak{D}_i$ is compatible with $\phi_{ij}$.
It is naturally understood that indices of quantities ``in the $g_{\mu\nu}$ sector'' are raised or lowered with $\gamma^{ij}$ and $\gamma_{ij}$, whereas $\phi^{ij}$ and $\phi_{ij}$ act in those ``in the $f_{\mu\nu}$ sector'' (the two sectors are well enough separated so this will not be confusing). 

A Legendre transformation yields the precursor primary Hamiltonian
\begin{equation}\label{eq:precursor_H}
H_\text{pre}^{(1)} = -\int\!\dd^3x\left(N\mathcal{R}_0+N^i\mathcal{R}_i+M\tilde{\mathcal{R}}_0+M^i\tilde{\mathcal{R}}_i\right).
\end{equation}
As in GR, the Hamiltonian is a linear combination of constraints and thus vanishes on shell up to a boundary term, which does not affect the number of local degrees of freedom and thus can be safely dropped for our purpose. Those constraints read
\begin{subequations}\label{eq:prec_contr}
\begin{align}
& \mathcal{R}_0 = \mathcal{R}_0^\text{GR}- \frac{m^2\Mpl^2}{2}\,\sqrt\gamma\,\mathcal{H}_0\,,
& & \tilde{\mathcal{R}}_0 = \tilde{\mathcal{R}}_0^\text{GR}- \frac{m^2\Mpl^2}{2}\,\sqrt\phi\,\tilde{\mathcal{H}}_0\,, \\
& \mathcal{R}_i = 2\sqrt\gamma\gamma_{ij}\mathcal{D}_k\left(\frac{\pi^{jk}}{\sqrt\gamma}\right)\,,
& & \tilde{\mathcal{R}}_i = 2\sqrt\phi\phi_{ij}\mathfrak{D}_k\left(\frac{\sigma^{jk}}{\sqrt\phi}\right)\,,
\\
& \mathcal{R}_0^\text{GR} = \frac{\Mpl^2}{2}\sqrt\gamma\,R\left[\gamma\right]-\frac{2}{\Mpl^2\sqrt\gamma}\left(\pi^{pq}\pi_{pq}-\frac{\pi^2}{2}\right)\,,
\quad
& & \tilde{\mathcal{R}}_0^\text{GR} = \frac{\alpha^2\Mpl^2}{2}\sqrt\phi\,R\left[\phi\right]-\frac{2}{\alpha^2\Mpl^2\sqrt\phi}\left(\sigma^{pq}\sigma_{pq}-\frac{\sigma^2}{2}\right)\,,
\\
& \mathcal{H}_0 = \sum_{n=0}^3c_{4-n}e_n\left(\mathfrak{K}\right)\,,
& & \tilde{\mathcal{H}}_0 = \sum_{n=0}^3c_{n}e_n\left(\mathcal{K}\right)\,,
\end{align}
\end{subequations}
with the natural shorthands for the traces $\pi = \gamma_{ij}\pi^{ij}$ and $\sigma= \phi_{ij}\sigma^{ij}$. 

In order to count the number of \emph{dof} that are present at this stage, we need to discriminate first-class constraints (i.e.\ those whose Poisson brackets with all constraints vanish weakly and each of which removes two phase space \emph{dof} at each point) from second-class ones (i.e.\ those that are not first-class and each of which only removes one phase space \emph{dof} at each point). 
Therefore, one has to compute the algebra of Poisson brackets among all independent constraints. 
For convenience, we will express those Poisson brackets using ``distributional forms'' of the constraints, i.e.\ using well-behaved test-functions $\zeta$ and $\xi^i$, that are assumed to vanish on the spatial boundary, as
\begin{equation}
\mathcal{R}_0 \mapsto \mathcal{R}_0\left[\zeta\right] \equiv \int\!\dd^3y \,\mathcal{R}_0 \,\zeta(y)\,,
\qquad
\mathcal{R}_i \mapsto \mathcal{R}_i\left[\xi^i\right] \equiv \int\!\dd^3y \,\mathcal{R}_i \,\xi^i(y) = - 2\int\!\dd^3y \,\pi^j_i\mathcal{D}_j \,\xi^i\,,
\nonumber
\end{equation}
and similarly in the $f$-sector, and denote with the weak equality ``$\approx$'' on-shell relations.

\renewcommand{\arraystretch}{1.5}
\begin{table}[h!]
\begin{center}
\begin{tabular}{|c||c|c|c|c|}
\hline
$\left\lbrace \downarrow , \rightarrow \right\rbrace$ & $\mathcal{R}_0$ & $\mathcal{R}_i$ & $\tilde{\mathcal{R}}_0$ & $\tilde{\mathcal{R}}_i$ \\
\hline
\hline
$\mathcal{R}_0$ & $\approx 0$ & $\mathcal{A}_i$ & $\mathcal{B}_0$ & $\tilde{\mathcal{B}}_i$\\
\hline
$\mathcal{R}_j$ &  & $\approx 0$ & $-\mathcal{B}_i$ & $0$\\
\hline
$\tilde{\mathcal{R}}_0$ &  &  & $\approx 0$ & $\tilde{\mathcal{A}}_i$\\
\hline
$\tilde{\mathcal{R}}_j$ &  & & & $\approx 0$\\
\hline
\end{tabular}
\end{center}
\caption{Constraint algebra of the precursor theory. The omitted entries are due to the antisymmetric nature of the Poisson brackets.}
\label{table:precursor}
\end{table}

The detailed computation of the Poisson brackets is presented in App.~\ref{app:Ham_pre}. The algebra of constraints is presented in Table.~\ref{table:precursor}, for which we have introduced the following notation, 
\begin{subequations}
\begin{align}
& \mathcal{A}_i\left[\xi^i\right] =\frac{m^2\Mpl^2}{4}\int\!\dd^3y\sqrt\gamma
\sum_{n=1}^3c_{5-n}\left( U_{(n)\, j}^i +\gamma^{iq}\gamma_{jp} \, U_{(n)\, q}^p
\right)\mathcal{D}_i\xi^j
-\frac{m^2\Mpl^2}{2}\int\!\dd^3y\sqrt\gamma\mathcal{D}_i\left(\mathcal{H}_0\xi^i\right)\,,\\
& \tilde{\mathcal{A}}_i\left[\xi^i\right] =\frac{m^2\Mpl^2}{4}\int\!\dd^3y\sqrt\phi
\sum_{n=1}^3c_{n-1}\left( V_{(n)\, j}^i +\phi^{iq}\phi_{jp} \, V_{(n)\, q}^p \right)\mathfrak{D}_i\xi^j
-\frac{m^2\Mpl^2}{2}\int\!\dd^3y\sqrt\phi\mathfrak{D}_i\left(\tilde{\mathcal{H}}_0\xi^i\right)\,,\\
& \mathcal{B}_0 = - m^2
\sum_{n=1}^4\left[c_{4-n}\left(\pi^i_j-\frac{\pi}{2}\delta^i_j\right)U^j_{(n)\,i}-\frac{c_n}{\alpha^2}\left(\sigma^i_j-\frac{\sigma}{2}\delta^i_j\right)V^j_{(n)\,i}\right]\,,\\
& \mathcal{B}_i\left[\xi^i\right] =\frac{m^2\Mpl^2}{4}\int\!\dd^3y\sqrt\gamma
\sum_{n=1}^3c_{4-n}\left( U_{(n)\, j}^i +\gamma^{iq}\gamma_{jp} \, U_{(n)\, q}^p
\right)\mathcal{D}_i\xi^j\,,\\
& \tilde{\mathcal{B}}_i\left[\xi^i\right] =\frac{m^2\Mpl^2}{4}\int\!\dd^3y\sqrt\phi
\sum_{n=1}^3c_n\left( V_{(n)\, j}^i +\phi^{iq}\phi_{jp} \, V_{(n)\, q}^p \right)\mathfrak{D}_i\xi^j\,,
\end{align}
\end{subequations}
where the derivatives of the elementary symmetric polynomials are given by
\begin{equation}
U_{(n)\, q}^p \equiv \frac{\partial e_n(\mathfrak{K})}{\partial \mathfrak{K}^q_{\ p}} = \sum_{\ell =0}^{n-1} (-)^\ell e_{n-1-\ell}(\mathfrak{K})\, \left(\mathfrak{K}^\ell\right)^p_{\ q}\,,
\end{equation}
and similarly for $V_{(n)\, q}^p \equiv \frac{\partial e_n(\mathcal{K})}{\partial \mathcal{K}^q_{\ p}}$.

The matrix in Table~\ref{table:precursor} shows the Poisson brackets among constraints with the delta function omitted and is of rank four. There are thus four first-class and four second-class constraints at each point at this level. Note that among the first-class constraints are the three combinations $\mathcal{R}_i + \tilde{\mathcal{R}}_i$, as will be proven in App.~\ref{app:Ham_MTBG}. (Actually, as the precursor theory is promoted to MTBG by additional constraints in the next section, the three combinations $\mathcal{R}_i + \tilde{\mathcal{R}}_i$ remain first-class.) 
All those constraints remove a total of twelve phase space \emph{dof} at each point. We thus have at most twelve phase space \emph{dof} left at each point, so we still have to eliminate four of them, by introducing appropriate constraints.

\section{The Minimal Theory of Bigravity}
\label{sec:MTBG}

\subsection{Hamiltonian formulation}
\label{subsec:MTBG-hamiltonian}

We define the Minimal Theory of Bigravity (MTBG) by adding constraints to the precursor Hamiltonian, such that the theory contains only four gravitational \emph{dof}. This is realized by the Hamiltonian
\begin{equation}
\begin{aligned}
H  = -\int\!\dd^3x & \Biggl\lbrace N\mathcal{R}_0 + N^i\mathcal{R}_i+M\tilde{\mathcal{R}}_0 + M^i\tilde{\mathcal{R}}_i + \lambda\left(\mathcal{C}_0-\tilde{\mathcal{C}}_0\right) + \lambda^i\left(\mathcal{C}_i- \beta\,\tilde{\mathcal{C}}_i\right) \\
& \left.\qquad + \bar\lambda\left[ \sqrt\gamma \gamma^{ij}\mathcal{D}_{ij}\left(\frac{\mathcal{C}_0}{\sqrt\gamma}\right)+\sqrt\phi \phi^{ij}\mathfrak{D}_{ij}\left(\frac{\tilde{\mathcal{C}}_0}{\sqrt\phi}\right)\right] \right\rbrace\,, \label{eq:MTBG_H}
\end{aligned}
\end{equation}
where $\{\lambda,\bar\lambda,\lambda^i\}$ is a set of five Lagrange multipliers, $\beta$ is a yet free constant and the eight first constraints $\{\mathcal{R}_0,\mathcal{R}_i,\tilde{\mathcal{R}}_0,\tilde{\mathcal{R}}_i\}$ are the same as in Eq.~\eqref{eq:prec_contr}. The remaining pieces are defined in a similar way as what was done when constructing MTMG~\cite{MTMG_OP}, namely
\begin{subequations}
\begin{align}
& \mathcal{C}_0\left[\zeta\right] = 
\left\lbrace \mathcal{R}_0^\text{GR}\left[\zeta\right],-\frac{m^2\Mpl^2}{2}\,\int\!\text{d}^3x\,\sqrt\phi \tilde{\mathcal{H}}_0\right\rbrace\,,
&
& \mathcal{C}_i\left[\xi^i\right]=
\left\lbrace \mathcal{R}_i\left[\xi^i\right],-\frac{m^2\Mpl^2}{2}\,\int\!\text{d}^3x\,\sqrt\phi \tilde{\mathcal{H}}_0\right\rbrace \,,\label{eq:def_C0_Ci}\\
& \tilde{\mathcal{C}}_0\left[\zeta\right] = 
\left\lbrace \tilde{\mathcal{R}}_0^\text{GR}\left[\zeta\right],-\frac{m^2\Mpl^2}{2}\,\int\!\text{d}^3x\,\sqrt\gamma\mathcal{H}_0\right\rbrace\,,
&
& 
\tilde{\mathcal{C}}_i\left[\xi^i\right]=
\left\lbrace \tilde{\mathcal{R}}_i\left[\xi^i\right],-\frac{m^2\Mpl^2}{2}\,\int\!\text{d}^3x\,\sqrt\gamma\mathcal{H}_0\right\rbrace \,,
\end{align}
\end{subequations}
and read explicitly
\begin{subequations}
\begin{align}
& \mathcal{C}_0 = 
- m^2 \left(\pi^{pr}\gamma_{rq}-\frac{\pi}{2}\,\delta^p_q\right)\mathcal{U}^q_{\ p}\,,
&
& \mathcal{C}_i=
\frac{m^2\Mpl^2}{2}\,\sqrt\gamma
\,\mathcal{D}_j \mathcal{U}^j_{\ i}\,,\\
& \tilde{\mathcal{C}}_0=
- \frac{m^2}{\alpha^2} \left(\sigma^{pr}\phi_{rq}-\frac{\sigma}{2}\,\delta^p_q\right)\tilde{\mathcal{U}}^q_{\ p}\,,
&
& 
\tilde{\mathcal{C}}_i=
\frac{m^2\Mpl^2}{2}\,\sqrt\phi
\,\mathfrak{D}_j \tilde{\mathcal{U}}^j_{\ i}\,,
\end{align}
\end{subequations}
with the shorthands
\begin{equation}
\mathcal{U}^p_{\ q} \equiv \frac{1}{2}\sum_{n=1}^3c_{4-n}\left(U^p_{(n)\,q}+\gamma^{pr}\gamma_{qs}U^s_{(n)\,r}\right)\,,
\quad \text{and} \qquad
\tilde{\mathcal{U}}^p_{\ q} \equiv \frac{1}{2}\sum_{n=1}^3c_n\left(V^p_{(n)\,q}+\phi^{pr}\phi_{qs}V^s_{(n)\,r}\right)\,.
\end{equation}
The formulation of the theory in the vielbein language is presented in App.~\ref{app:Vielbein}.
It may seem that we have added five more constraints with respect to the precursor Hamiltonian~\eqref{eq:precursor_H}, and thus that we would eliminate too many \emph{dof}.
Nevertheless, as we will see hereafter, one of the four first-class constraints of the precursor theory is demoted to second-class, thus we have added the accurate number of constraints.

We want here to discuss the reason why we made this choice for the extra constraints. First of all, the choice of the $\mathcal{C}_0 - \tilde{\mathcal{C}}_0$ constraint (i.e.\ the $\lambda$ constraint) was motivated as to give MTBG the same background equations of motion of HRBG on a generic FLRW manifold. In this case, MTBG naturally becomes a stable (its stability will be proven later on) nonlinear completion of the HRBG theory. Having now, as one of the main goals of MTBG, a stable FLRW background makes the cosmology of the HRBG-background appealing once again.
Since, as already stated above, we need to add another second-class constraint to the theory, in order not to spoil the HRBG-background obtained by setting the $\lambda$-constraint, we have to introduce a 3D-scalar combination which does not affect the FLRW background in any possible way. It is then clear why we have introduced a constraint in the form of the one set by the Lagrange multiplier $\bar\lambda$. In fact, such a constraint indeed has no influence whatsoever on the FLRW background dynamics, however, it does affect, in general, the dynamics of other background solutions and, at least, the linear perturbations around FLRW manifolds. We recognize here that possible choices for other constraints (especially replacing the $\bar\lambda$ one) are in principle possible, but we will not discuss these possibilities in the present work. Needless to say, each inequivalent choice of constraints would lead to a different theory, in general. Finally, as for the $\lambda^i$ constraints, their number (three) comes as a consequence of 3D invariance, which is not broken in any way in MTBG. However, just by fixing to three the number of these vector-like constraints, several other possibilities could be taken into account. We have made an attempt to treat a rather general case, by introducing the parameter $\beta$ as an extra possible free parameter of the theory.
However, this $\beta$-constant will play no role in the cosmological study of Sec.~\ref{sec:cosmo}, and thus its numerical value is not relevant for the present work.
Actually, although we were tempted to fix it to unity by symmetry with the $\mathcal{C}_0 - \tilde{\mathcal{C}}_0$ constraint, we have chosen to let it be free, as it may play a role (and so be constrained) in the study of strong gravity regime (e.g.\ for black-hole, non-FLRW, type solutions).

We shall show that the theory~\eqref{eq:MTBG_H} contains at most four degrees of freedom. As we started with twenty-four phase space variables at each point and our theory contains thirteen constraints, it suffices to show that three of those constraints are first-class. 
In fact, although none of the constraints in~\eqref{eq:MTBG_H} is explicitly first-class, we show in the App.~\ref{app:Ham_MTBG} that the combinations
\begin{equation}\label{eq:def_Ri_3D}
\mathrm{R}_i \equiv \mathcal{R}_i+\tilde{\mathcal{R}}_i\,,
\end{equation}
are still the first-class constraints that generate spatial diffeomorphism. This is similar to what is happening in usual bigravity~\cite{Yamashita}.
In order to demonstrate that this theory propagates exactly four physical \emph{dof} at each point, we should in principle properly check that the algebra of the Poisson brackets is closed. As this seems to be extremely involving, we will do a little detour and investigate the cosmology in Sec.~\ref{sec:cosmo}. This will allow us to prove that this theory propagates at least four \emph{dof}, thus we will conclude that it propagates exactly four \emph{dof}. Finally, as shown in App.~\ref{app:Ham_MTBG}, the fact that $\mathrm{R}_i$ are first-class constraints is totally independent of the value of the constant $\beta$.

\subsection{Lagrangian formulation}

From the Hamiltonian \eqref{eq:MTBG_H}, it follows that 
\begin{subequations}
\begin{align}
& \partial_t\gamma_{ij} = \frac{\partial H}{\partial \pi^{ij}}
= \frac{4N}{\Mpl^2\sqrt\gamma}\left(\pi_{ij}-\frac{\pi}{2}\,\gamma_{ij}\right) + 2\gamma_{k(i}\mathcal{D}_{j)}N^k
+m^2 \left(\mathcal{U}^p_{\ (i}\gamma_{j)p}-\frac{1}{2}\,\mathcal{U}^p_{\ p}\gamma_{ij}\right)\left(\lambda + \gamma^{kl}\mathcal{D}_{kl}\bar\lambda\right)\,,\\
& 
\partial_t\phi_{ij} =\frac{\partial H}{\partial \sigma^{ij}}
= \frac{4M}{\alpha^2\Mpl^2\sqrt\phi}\left(\sigma_{ij}-\frac{\sigma}{2}\,\phi_{ij}\right) + 2\phi_{k(i}\mathfrak{D}_{j)}M^k
-\frac{m^2}{\alpha^2}\left(\tilde{\mathcal{U}}^p_{\ (i}\phi_{j)p}-\frac{1}{2}\,\tilde{\mathcal{U}}^p_{\ p}\phi_{ij}\right)\left(\lambda - \phi^{kl}\mathfrak{D}_{kl}\bar\lambda\right)\,,
\end{align}
\end{subequations}
which can be inverted to give
\begin{subequations}
\begin{align}
& \pi^{ij} 
= \frac{\Mpl^2}{2}\sqrt\gamma\left\lbrace
K^{ij}-K\,\gamma^{ij}-\frac{m^2\left(\lambda + \gamma^{kl}\mathcal{D}_{kl}\bar\lambda\right)}{2N}\,\mathcal{U}^{(i}_{\ p}\gamma^{j)p}\right\rbrace\,,\\
& \sigma^{ij} 
= \frac{\alpha^2\Mpl^2}{2}\sqrt\phi\left\lbrace
\Phi^{ij}-\Phi\,\phi^{ij}
+\frac{m^2\left(\lambda - \phi^{kl}\mathfrak{D}_{kl}\bar\lambda\right)}{2M\alpha^2}\,\tilde{\mathcal{U}}^{(i}_{\ p}\phi^{j)p}\right\rbrace\,.
\end{align}
\end{subequations}
Legendre-transforming the Hamiltonian, the Lagrangian density of MTBG finally reads
\begin{equation}\label{eq:MTBG_L}
\begin{aligned}
 \mathcal{L} = & \:
\frac{\Mpl^2}{2}\,\sqrt{-g}\,\mathcal{R}\left[g\right]
+ \frac{\alpha^2\Mpl^2}{2}\,\sqrt{-f}\,\mathcal{R}\left[f\right]
-\frac{m^2\Mpl^2}{2}\left(
N\sqrt\gamma\,\mathcal{H}_0
+ M\sqrt\phi\,\tilde{\mathcal{H}}_0\right)\\
& 
- \left.\frac{m^2\Mpl^2}{2}\right\lbrace
\sqrt\gamma\,\mathcal{U}^p_{\ q}\mathcal{D}_p\lambda^q
- \beta \sqrt\phi\,\tilde{\mathcal{U}}^p_{\ q}\mathfrak{D}_p\lambda^q
+ \left(\lambda + \gamma^{ij}\mathcal{D}_{ij}\bar\lambda\right)\sqrt\gamma\,\mathcal{U}^p_{\ q}K^p_{\ q}
- \left(\lambda - \phi^{ij}\mathfrak{D}_{ij}\bar\lambda\right)\sqrt\phi\,\tilde{\mathcal{U}}^p_{\ q}\Phi^p_{\ q}\\
&
\qquad\qquad\qquad \left.
+ \frac{m^2\left(\lambda + \gamma^{ij}\mathcal{D}_{ij}\bar\lambda\right)^2}{4N}\sqrt\gamma\left(\mathcal{U}^p_{\ q}-\frac{\mathcal{U}^k_{\ k}}{2}\,\delta^p_q\right)\mathcal{U}^q_{\ p}
+ \frac{m^2\left(\lambda - \gamma^{ij}\mathfrak{D}_{ij}\bar\lambda\right)^2}{4M\alpha^2}\sqrt\phi\left(\tilde{\mathcal{U}}^p_{\ q}-\frac{\tilde{\mathcal{U}}^k_{\ k}}{2}\,\delta^p_q\right)\tilde{\mathcal{U}}^q_{\ p}
\right\rbrace\,.
\end{aligned}
\end{equation}
We are now ready to discuss the cosmology of MTBG.

\section{Cosmology}
\label{sec:cosmo}

In this section we present the cosmological solutions of the theory~\eqref{eq:MTBG_L}, at the background and linear perturbation levels.

\subsection{Cosmological background}

Let us first consider homogeneous and isotropic background solutions, i.e.\ let us begin with two flat Friedman-Lema\^{i}tre-Robertson-Walker (FLRW) line elements
\begin{equation}
\begin{aligned}
& \dd s_g^2 = -N^2(t)\,\dd t^2+a^2(t)\,\delta_{ij}\,\dd x^i\dd x^j\,,\\
& \dd s_f^2 = -M^2(t)\,\dd t^2+b^2(t)\,\delta_{ij}\,\dd x^i\dd x^j\,,\\
\end{aligned}
\end{equation}
and define the two Hubble-Lema\^{i}tre expansion rates $H = \dot{a}/Na$ and $L = \dot{b}/Mb$.
We will also introduce the ratios of the scale factors and the ratio of speeds of light in each sector as 
\begin{equation}
\mathcal{X} \equiv \frac{b}{a}\,,
\qquad\text{and}\qquad
r \equiv \frac{M}{N\mathcal{X}} = \frac{c_f}{c_g}\,.
\end{equation}
By virtue of homogeneity of the background, we will consider only time-dependent auxiliary fields $\lambda(t)$, $\bar\lambda(t)$ and $\lambda^i(t)$, so that $\bar\lambda$ and $\lambda^i$ disappear from the action.\footnote{The fact that $\lambda^i$ disappears from the action is also naturally enforced by the joint requirement of isotropy.}
As for the matter, we will consider two general perfect fluids with barotropic equations of state $p(\rho)$ and $\tilde{p}\left(\tilde{\rho}\right)$, minimally coupled respectively to $g_{\mu\nu}$ and $f_{\mu\nu}$.

With the mini-superspace ansatz, the gravitational action in a comoving volume $V$ reads
\begin{equation}
\mathcal{S} = \frac{\Mpl^2\,V}{2}\int\!\dd tNa^3\left\lbrace
-6H^2-6\frac{M\mathcal{X}^3\alpha^2}{N}\,L^2
-m^2\Gamma_0-\frac{m^2M}{N}\,\tilde{\Gamma}_0\\
-3m^2\left(H-\mathcal{X} L\right)\Gamma_1\lambda
-\frac{3m^4\Gamma_1^2\lambda^2}{8\alpha^2N^2\mathcal{X}^2}\left(\frac{1}{r}+\mathcal{X}^2\alpha^2\right)
\right\rbrace\,,\label{eq:lag_mini}
\end{equation}
where we have introduced
\begin{subequations}
\begin{align}
& \Gamma_0 = c_1\mathcal{X}^3+3c_2\mathcal{X}^2+3c_3\mathcal{X}+c_4\,,\\
& \tilde{\Gamma}_0 = c_0\mathcal{X}^3+3c_1\mathcal{X}^2+3c_2\mathcal{X}+c_3\,,\\
& \Gamma_1 = c_1\mathcal{X}^2+2c_2\mathcal{X}+c_3\,.
\end{align}
\end{subequations}

The equations of motion $\mathcal{E}_*= 0$ for the components of the metrics and the auxiliary field are respectively
\begin{subequations}\label{eq_cosmo_eom_grav}
\begin{align}
& \mathcal{E}_N = 3H^2-\frac{\rho+\rho_m+\rho_\text{aux}}{\Mpl^2}\,, \label{eq_cosmo_eom_N}\\
& \mathcal{E}_M = 3L^2-\frac{\tilde\rho+\tilde\rho_m+\tilde\rho_\text{aux}}{\alpha^2\Mpl^2}\,, \label{eq_cosmo_eom_M}\\
& \mathcal{E}_a = \frac{2\dot{H}}{N}+3H^2+\frac{p + p_m+p_\text{aux}}{\Mpl^2}\,,\label{eq_cosmo_eom_a}\\
& \mathcal{E}_b = \frac{2\dot{L}}{M}+3L^2+\frac{\tilde{p}+\tilde{p}_m+\tilde{p}_\text{aux}}{\alpha^2\Mpl^2}\,,\label{eq_cosmo_eom_b}\\
& \mathcal{E}_\lambda = 
\left(\frac{\Gamma_1\lambda}{N}+ \frac{\Gamma_1\lambda}{\alpha^2M\mathcal{X}}+\frac{4(H-\mathcal{X}L)}{m^2}\right)\Gamma_1\,,
\label{eq_cosmo_eom_lambda}
\end{align}
\end{subequations}
where we have introduced the usual matter energy densities and pressures $\rho$, $\tilde\rho$, $p$ and $\tilde{p}$, together with the effective energy densities
\begin{subequations}
\begin{align}
& \rho_m =\frac{\Mpl^2m^2\Gamma_0}{2} \,,
& & \tilde\rho_m = \frac{\Mpl^2m^2\tilde{\Gamma}_0}{2\mathcal{X}^3}\,,\label{eq:cosmo_rho_m}\\
& \rho_\text{aux} = -\frac{3m^2\Mpl^2\lambda}{2N}\left(H+\frac{m^2\Gamma_1\lambda}{8N}\right)\Gamma_1\,,
&& \tilde{\rho}_\text{aux} = \frac{3m^2\Mpl^2\lambda}{2M\mathcal{X}^2}\left(L-\frac{m^2\Gamma_1\lambda}{8M\alpha^2\mathcal{X}^2}\right)\Gamma_1\,,
\end{align}
\end{subequations}
and pressures
\begin{subequations}
\begin{align}
& p_m = - m^2\Mpl^2\,\frac{\Gamma_0+\mathcal{X}\Gamma_1 \left(r-1\right)}{2}\,,
\qquad 
\tilde{p}_m = - \frac{m^2\Mpl^2}{2\mathcal{X}^3}\left(\tilde{\Gamma}_0+\frac{1-r}{r}\Gamma_1\right)\,, \label{eq:cosmo_p_m}\\
&
p_\text{aux} = \frac{m^2\Mpl^2}{2N}\left\lbrace
\frac{\dd}{\dd t}\left[\frac{\Gamma_1\lambda}{N}\right] - 2\left[\left(c_1\mathcal{X}+c_2\right)H+\left(c_2\mathcal{X}+c_3\right)L\right]\mathcal{X}\lambda \right.\\
& \hspace{7cm}\left.
-\frac{m^2\lambda^2\Gamma_1\mathcal{X}}{2M}\left[\frac{c_2\mathcal{X}+c_3}{\alpha^2\mathcal{X}^2}-\frac{c_1\mathcal{X}^2-2c_2\mathcal{X}-3c_3}{4}\,r\right]\right\rbrace\,,\nonumber\\
&
\tilde{p}_\text{aux} = -\frac{m^2\Mpl^2}{2M}\left\lbrace
\frac{\dd}{\dd t}\left[\frac{\Gamma_1\lambda}{M\mathcal{X}^2}\right]
+ 2 \left[ \left(c_1\mathcal{X}+c_2\right)H+\left(c_2\mathcal{X} +c_3\right)L\right]\frac{\lambda}{\mathcal{X}^2}\right. \\
& \hspace{7cm}\left.
+\frac{m^2\lambda^2\Gamma_1}{2N\mathcal{X}^2}\left[c_1\mathcal{X}+c_2+\frac{3c_1\mathcal{X}^2+2c_2\mathcal{X}-c_3}{4\alpha^2\mathcal{X}\,r}\right]\right\rbrace\,.\nonumber
\end{align}
\end{subequations}
The two perfect fluids obey their respective continuity equations
\begin{equation}
\mathcal{E}_c^g = \dot\rho +3NH\left(\rho+p\right)\,,
\qquad \text{and} \qquad
\mathcal{E}_c^f = \dot{\tilde\rho} +3ML\left(\tilde\rho+\tilde{p}\right)\,.
\label{eq_cosmo_eom_mat}
\end{equation}
In addition, let us also define two ``Bianchi'' identities, one for each sector, that will be used later on 
\begin{equation}\label{eq:Bianchi_bckg}
\mathcal{E}_B^g = \dot{\mathcal{E}}_N + 3NH\left(\mathcal{E}_N-\mathcal{E}_a\right) + \frac{\mathcal{E}_c^g}{\Mpl^2}\,,
\qquad \text{and}\qquad
\mathcal{E}_B^f = \dot{\mathcal{E}}_M + 3ML\left(\mathcal{E}_M-\mathcal{E}_b\right) + \frac{\mathcal{E}_c^f}{\alpha^2\Mpl^2}\,.
\end{equation}

The equations of motion for the metric~\eqref{eq_cosmo_eom_N}-\eqref{eq_cosmo_eom_b} and the continuity equations~\eqref{eq_cosmo_eom_mat} take the same form as in the usual HRBG~\cite{bigrav_viable_cosmo} (for the reader's convenience, a dictionary between our notations and those of~\cite{bigrav_viable_cosmo} is provided in App.~\ref{app:dic}).
Moreover the equation of motion for the auxiliary field $\lambda$~\eqref{eq_cosmo_eom_lambda} reveals that there exist two branches of solutions, that are the same as the two branches of HRBG cosmology. 
As we will prove hereafter, the ``normal'' branch, defined by $H=\mathcal{X} L$ corresponds to the healthy branch of~\cite{bigrav_viable_cosmo}.
The second, ``self-accelerating'' branch, defined by $\Gamma_1 = 0$ (and thus a constant $\mathcal{X}$) is the analogue of the strongly-coupled one of bigravity\footnote{The self-accelerating and normal branches have also been called ``algebraic'' and ``dynamical'', respectively (for example in \cite{bigrav_Planck_hierarchy}).}, up to the only important difference that it is now perfectly healthy in MTBG. For a discussion of the background equations of motion in the Hamiltonian formalism see App.~\ref{app:hamil_back}.

\subsubsection{Self-accelerating branch}

In the self-accelerating branch, $\Gamma_1 = 0$ and thus $\mathcal{X}$ is constant, which implies in turn $ML = NH$.
Closing the set of equations of motion~\eqref{eq_cosmo_eom_grav} and \eqref{eq_cosmo_eom_mat} is equivalent to verify that the identities~\eqref{eq:Bianchi_bckg} hold. They are proportional to each other and give the constraint
\begin{equation}
3m^2 H^2\left(c_1\mathcal{X}+c_2\right)\frac{1-\mathcal{X}^2\,r}{\mathcal{X}\,r}\,\lambda =0\,.
\end{equation}
As we will see later, in addition to be a severe tuning, strong coupling issues appear when the quantity $c_1\mathcal{X}+c_2$ is vanishing, so we will keep it non-zero. On the other hand, imposing $\mathcal{X}^2\,r =1$ would yield $H = L/\mathcal{X}$ which, as $\mathcal{X}$ is constant, totally fixes the dynamics of the $g$-sector with respect to the $f$-sector. This would naturally yield an ill-defined cosmology, thus we discard this possibility. Therefore the only possible way to satisfy the Bianchi's identities is to fix $\lambda = 0$.

Defining the two constants
\begin{equation}
\Lambda_0 = \frac{m^2}{2}\left(c_2\mathcal{X}^2+2c_3\mathcal{X}+c_4\right)
\qquad \text{and} \qquad
\tilde{\Lambda}_0 = \frac{m^2}{2\alpha^2}\frac{c_0\mathcal{X}^2+2c_1\mathcal{X}+c_2}{\mathcal{X}^2}\,,
\end{equation}
the energies and pressures coming from the mass sector are then
\begin{equation}
\rho_m =\Mpl^2\,\Lambda_0 \,,
\qquad p_m = -\rho_m\,,
\qquad \tilde{\rho}_m =\alpha^2\Mpl^2\,\tilde{\Lambda}_0 \,,
\qquad \tilde{p}_m = -\tilde{\rho}_m\,
\end{equation}
so that the mass sector yields a set of two effective cosmological constants, thus the name ``self-accelerating'' branch. As we will show later on, since MTBG has only tensor modes, this branch is not strongly coupled any longer, and one can then study its phenomenology.

\subsubsection{Normal branch}

In the normal branch, recalling that by definition $\Gamma_1 \neq 0$, eq.~\eqref{eq_cosmo_eom_lambda} is solved by 
\begin{equation}
\lambda = \frac{4\alpha^2N\mathcal{X}^2\,r\,\left(\mathcal{X}L-H\right)}{m^2\Gamma_1\left(1+\alpha^2\mathcal{X}^2\,r\right)}\,.
\end{equation}
Injecting this result in the identities~\eqref{eq:Bianchi_bckg} and forming the linear combination $\mathcal{E}_B^g + \mathcal{X}^4\, \mathcal{E}_B^f$ yields
\begin{equation}
\mathcal{E}_B^g + \mathcal{X}^4\, \mathcal{E}_B^f = E\left(H-\mathcal{X}L\right) = 0\,,
\end{equation}
where $E$ is a complicated expression that has no reason to vanish (or it will be equivalent to add more constraints). This relation thus imposes that $H-\mathcal{X}L$, and thus $\lambda$, vanish.
So in the normal branch, we have 
\begin{equation}
\lambda = 0
\qquad \text{and} \qquad
H = \mathcal{X}L\,.
\end{equation}
Note here that the previous argument against proportionality between $H$ and $L$ does not hold here as neither $\mathcal{X}$ nor $r$ are constant.
The normal sector is thus filled with two effective dark fluids ($\rho_m$, $p_m$) and ($\tilde\rho_m$, $\tilde{p}_m$) that are specified by \eqref{eq:cosmo_rho_m} and \eqref{eq:cosmo_p_m} and that satisfy conservation equations. 
The equations of state of those effective dark fluids are given by
\begin{subequations}\label{eq:eos_Normbranch}
\begin{align}
& w_m \equiv \frac{p_m}{\rho_m} = -1 + \frac{c_1\mathcal{X}^3 + 2 c_2\mathcal{X}^2 + c_3\mathcal{X}}{c_1\mathcal{X}^3 +3 c_2\mathcal{X}^2 + 3c_3\mathcal{X}+ c_4}\,\left(1-r\right)\,,\\
& \tilde{w}_m \equiv \frac{\tilde{p}_m}{\tilde\rho_m} = -1 + \frac{c_1\mathcal{X}^3 + 2 c_2\mathcal{X}^2 + c_3\mathcal{X}}{c_0\mathcal{X}^3 +3 c_1\mathcal{X}^2 + 3c_2\mathcal{X}+ c_3}\,\frac{r-1}{r}\,.
\end{align}
\end{subequations}

This branch is analogous to the healthy branch of HRBG cosmology, defined by $H = \mathcal{X}L$~\cite{bigrav_viable_cosmo}. 
Moreover, as in HRBG, the evolution of the ratios $\mathcal{X}$ and $r$ is fixed by the matter content, as 
\begin{equation}\label{eq_cosmo_XR_evol}
\frac{\dot{\mathcal{X}}}{N\mathcal{X}} = \left(r-1\right)\,H\,,
\qquad\text{and}\qquad
2\left(\frac{m^2\Gamma_1}{4\mathcal{X}}\frac{1+\alpha^2\mathcal{X}^2}{\alpha^2}-H^2\right)\left(r-1\right) = \frac{\rho+p}{\Mpl^2}-\frac{\tilde\rho+\tilde{p}}{\alpha^2\Mpl^2}\,\mathcal{X}^2\,r\,.
\end{equation}

\subsection{Cosmological perturbations}

Let us now turn to the study of linear cosmological perturbations, on both branches.
This study has two aims: on one hand, by showing that 4 gravitational \emph{dof} propagates at the linear level, we can conclude that the theory contains exactly 4 gravitational \emph{dof} since the Hamiltonian analysis in subsection~\ref{subsec:MTBG-hamiltonian} has already shown that the number of propagating \emph{dof} is at most four at the fully nonlinear level. 
On the other hand, it allows us to investigate the cosmological phenomenology, and possible deviations from canonical GR results.

Let us perturb the gravitational sector as
\begin{subequations}
\begin{align}
& N = N(t)\left(1+ \eta\right)\,,
& \quad
& M = M(t)\left(1+ \omega\right)\,,\\
& N_i = N(t)\,a^2(t)\left(\beta_i+ \partial_i\beta\right)\,,
& 
& M_i = M(t)\,b^2(t)\left(\theta_i+ \partial_i\theta\right)\,,\\
& \gamma_{ij} = a^2(t)\left[\left(1+2\zeta\right)\delta_{ij} + 2 \partial_{ij}\chi+ 2\partial_{(i}\chi_{j)}+\pi_{ij}\right]\,,
& 
& \phi_{ij} = b^2(t)\left[\left(1+2\xi\right)\delta_{ij} + 2 \partial_{ij}\tau+ 2\partial_{(i}\tau_{j)}+\psi_{ij}\right]\,,\\
& \lambda = \delta\lambda\,,
&
& \bar\lambda = \bar\lambda(t) + \delta\bar\lambda\,,\\
& \lambda^i = \delta\ell^i+\frac{\delta^{ij}}{a^2(t)}\,\partial_j\delta\ell\,.
\end{align}
\end{subequations}
where all the perturbation variables depend on space and time, the vector quantities $\{\beta_i,\theta_i,\chi_i,\tau_i,\delta\ell^i\}$ are divergenceless and the tensor quantities $\{\pi_{ij},\psi_{ij}\}$ are divergenceless, symmetric and trace-free. We operate indices in both sectors with the flat $\delta_{ij}$ metric. 

As for the matter, we will present the computation with two pure dust components, \emph{\`{a} la} Schutz-Sorkin \cite{Schutz77,Brown93}, one in each sector
\begin{equation}
\mathcal{S}_\text{dust} = - \int\!\text{d}^4x\left\lbrace\sqrt{-g}\,\left[\rho(n) + J^\alpha\partial_\alpha\varphi\right]+\sqrt{-f}\,\left[\tilde{\rho}(\tilde{n}) + \tilde{J}^\alpha\partial_\alpha\tilde{\varphi}\right] \right\rbrace\,,
\end{equation}
where $J^\alpha$ and $\tilde{J}^\alpha$ are two four-vectors. The dusts are made of particles of individual masses $\mu_0$ and $\tilde{\mu}_0$ so that the densities read
\begin{subequations}\label{eq:def_rho_n_Schutz}
\begin{align}
& \rho = \mu_0\,n\,,
& & n = \sqrt{-g_{\alpha\beta}J^\alpha J^\beta}\,,\\
& \tilde\rho = \tilde{\mu}_0\,\tilde{n}\,,
& & \tilde{n} = \sqrt{-f_{\alpha\beta}\tilde{J}^\alpha \tilde{J}^\beta}\,,
\end{align}
\end{subequations}
and the four-velocities $u^\alpha = J^\alpha/n$ and $\tilde{u}^\alpha =\tilde{J}^\alpha/\tilde{n}$ are properly normalized $u^\alpha u_\alpha = \tilde{u}^\alpha\tilde{u}_\alpha = -1$. Varying $\mathcal{S}_\text{dust}$ \emph{w.r.t.}\ $J^\alpha$ and $\tilde{J}^\alpha$, it follows that
\begin{equation}
u_\alpha = \frac{\partial_\alpha\varphi}{\mu_0}\,,
\qquad\text{and}\qquad
\tilde{u}_\alpha = \frac{\partial_\alpha\tilde\varphi}{\tilde{\mu}_0}\,.
\end{equation}

This dust action is perturbed as
\begin{subequations}
\begin{align}
& J^0 = \frac{1}{Na^3}\left(\mathcal{N}_0+ j_0\right)\,,
& & \tilde{J}^0 = \frac{1}{Mb^3}\left(\tilde{\mathcal{N}}_0+\tilde{j}_0\right) \,, \label{eq:pert_J0} \\
& J^i = \frac{1}{a^3(t)}\left( j^i+\frac{\delta^{ik}}{a^2(t)}\,\partial_k j\right)\,,
& &  \tilde{J}^i =  \frac{1}{b^3(t)}\left(\tilde{j}^i+\frac{\delta^{ik}}{b^2(t)}\,\partial_k \tilde{j}\right)\,,\\
& \varphi = -\mu_0\int^t\text{d}\tau N(\tau)-\mu_0v_m\,,
& & \tilde\varphi = -\tilde{\mu}_0\int^t\text{d}\tau M(\tau)-\tilde{\mu}_0\tilde{v}_m\,,
\end{align}
\end{subequations}
where $\mathcal{N}_0$ and $\tilde{\mathcal{N}}_0$ are constants of integration of the background equations for $\varphi$ and $\tilde\varphi$, that represent the comoving number density of particles, as they satisfy $\rho = \mu_0\,\mathcal{N}_0 /a^3$ and  $\tilde\rho = \tilde{\mu}_0\,\tilde{\mathcal{N}}_0 /b^3$.
We will also introduce the gauge invariant density contrasts
\begin{equation}
\delta_m \equiv \frac{\delta\rho}{\rho} + 3H v_m\,,
\qquad\text{and}\qquad
\tilde{\delta}_m \equiv \frac{\delta\tilde\rho}{\tilde\rho} + 3L \tilde{v}_m\,.
\end{equation}
Using~\eqref{eq:def_rho_n_Schutz} and~\eqref{eq:pert_J0}, it follows that
\begin{equation}\label{eq:delta_m_j0}
\delta_m \equiv \frac{j_0}{\mathcal{N}_0} +\eta + 3H v_m\,,
\qquad\text{and}\qquad
\tilde{\delta}_m \equiv \frac{\tilde{j}_0}{\tilde{\mathcal{N}}_0} + \omega + 3L \tilde{v}_m\,.
\end{equation}

For a cross-check and as a complementary calculation, we have also made the computation with two minimally coupled $k$-essences
\begin{equation}\label{eq:Skess_def}
\mathcal{S}_{k-\text{ess}} = \int\!\text{d}^4x\left\lbrace\sqrt{-g}\,P(X)+\sqrt{-f}\,Q(Y)\right\rbrace\,,
\qquad\text{where}\qquad
X \equiv -\frac{g^{\mu\nu}\partial_\mu\phi\,\partial_\nu\phi}{2}\,,\quad
Y \equiv -\frac{f^{\mu\nu}\partial_\mu\,\tilde\phi\partial_\nu\tilde\phi}{2}\,.
\end{equation}
For this system of matter, the background pressures and energy densities are $p = P(X)$, $\tilde{p} = Q(Y)$, $\rho = 2XP_X-P$ and $\tilde\rho = 2YQ_Y-Q$.
As for the perturbations, the no-ghost conditions and squared sound speeds are given in the $g$-sector by
\begin{equation}\label{eq:Skess_noghost_cs2}
2XP_{XX}+P_X > 0\,,
\qquad\text{and}\qquad
c_s^2 = \frac{P_X}{2XP_{XX}+P_X}\,,
\end{equation}
with similar relations holding in the $f$-sector.

\subsubsection{Tensor sector}

In both branches, the quadratic action of the tensor sector reads
\begin{equation}\label{eq:tensor_pert}
\delta^{(2)}\,\mathcal{S}^T = 
\frac{\Mpl^2}{8}\int\!\dd t\dd^3xNa^3\left\lbrace 
\left(\frac{\dot{\pi}_{ij}}{N}\right)^2-
\left(\frac{\partial_k\pi_{ij}}{a}\right)^2
+\frac{\alpha^2\mathcal{X}^3M}{N}\left[\left(\frac{\dot{\psi}_{ij}}{M}\right)^2-
\left(\frac{\partial_k\psi_{ij}}{\mathcal{X}a}\right)^2\right]
- \mu_T^2\,\left(\pi_{ij}-\psi_{ij} \right)^2  \right\rbrace\,,
\end{equation}
the only difference being the expression of the effective mass
\begin{subequations}\label{eq:mu2_tensor}
\begin{align}
& \mu_T^2 = \frac{m^2\mathcal{X}^2}{2}\left(c_1\mathcal{X}+c_2\right)\left(r-1\right)& & \text{in the self-accelerating branch}\,,\\
& \mu_T^2 = \frac{m^2\mathcal{X}^2}{2}\,\left[\left(c_1\mathcal{X} + c_2\right)\,r+ \frac{c_2\mathcal{X}+c_3}{\mathcal{X}}\right] & & \text{in the normal branch}\,.
\end{align}
\end{subequations}

As the kinetic terms are the canonical ones, we are automatically free from ghost and gradient instabilities. In order to avoid tachyonic instabilities, one needs to ensure the positivity of the squared mass.
It is clear that the tensor sector propagates 4 \emph{dof} at linear level.
On the phenomenological point of view, we recover the usual bigravity features: the mass eigenstates are different from either $g$-perturbation or $f$-perturbation, with the massive eigenstate being $h_{ij} \propto \pi_{ij}-\psi_{ij}$. This leads to oscillations of the gravitons during their propagation.

\subsubsection{Vector sector}

In the vector sector, let us define $D_i = \chi_i - \tau_i$, $B_i = \beta_i - \dot{\chi}_i/N$ and similarly $\tilde{B}_i = \theta_i - \dot{\tau}_i/M$. In both branches, the quadratic action can then be represented as
\begin{equation}
\begin{aligned}
\delta^{(2)}\,\mathcal{S}^V = 
\frac{\Mpl^2}{4}\int\!\dd t\dd^3xNa^3 
& \left\lbrace 
\left(\partial_iB_j\right)^2 + \frac{\alpha^2\mathcal{X}^3M}{N}\left(\partial_i\tilde{B}_j\right)^2
-\mu_T^2\left(\partial_iD_j\right)^2
-m^2 \frac{\left(c_1\mathcal{X}+c_2\right)\left(1-\beta\mathcal{X}\right)-\beta\Gamma_1}{N}\partial_iD_j\,\partial^i\delta\ell^j
\right.\\
& \left.
+ \frac{2a^2\rho}{\Mpl^2}\left(\beta_i + \frac{j_i}{\mathcal{N}_0}\right)^2
+ \frac{2a^2\mathcal{X}^2\tilde{\rho}}{\Mpl^2\,r}\left(\theta_i + \frac{\tilde{j}_i}{\tilde{\mathcal{N}}_0}\right)^2
\right\rbrace\,,
\end{aligned}
\end{equation}
where $\mu_T^2$ is the squared tensor mass~\eqref{eq:mu2_tensor}.
The auxiliary field imposes the constraint $D_i =0$, which eliminates the whole massive sector, so the action simply reduces to two linked copies of the usual GR vector sector.
The phenomenology of the vector sector of MTBG is thus the same as the one of GR, separately in each metric.
Moreover this clearly shows that no vector gravitational degree of freedom is dynamical, as in GR.

\subsubsection{Scalar sector in the self-accelerating branch}

In the scalar sector of the self-accelerating branch, all auxiliary fields enter linearly in the quadratic action.
As variations \emph{w.r.t.}\ $\delta\lambda$ and $\delta\bar\lambda$ yield redundant constraints, one has only the two relations
\begin{equation}
\zeta = \xi\,,
\qquad\text{and}\qquad
\chi = \tau\,.
\end{equation}
Those two conditions suffice to eliminate the mass sector totally, leaving only two copies of the usual GR scalar sector.

In the following, we eliminate the quantities $j_0$ and $\tilde{j}_0$ in favor of the gauge invariant density contrasts~\eqref{eq:delta_m_j0}.
Moreover, we still have one freedom to fully fix the 3D gauge, and thus we will take $\zeta = 0$.
One can then integrate out $j$ and $\tilde{j}$ as
\begin{equation}\label{eq:deltaj_int_out}
j = -\mathcal{N}_0\left(v_m+a^2\beta\right)\,,
\qquad\text{and}\qquad
\tilde{j} = -\tilde{\mathcal{N}}_0\left(\tilde{v}_m+\mathcal{X}^2a^2\theta\right)\,,
\end{equation}
so that the shift perturbations $\beta$ and $\theta$ enter linearly and allow to express $\eta = \rho v_m/(2\Mpl^2H)$ and $\omega = \tilde\rho \tilde{v}_m/(2\alpha^2\Mpl^2L)$. At this point $\chi$ vanishes by virtues of its own equation of motion and $v_m$ and $\tilde{v}_m$ can be integrated out so to give
\begin{equation}
\delta^{(2)}\mathcal{S}^S_\text{sa} = \int\!\text{d}t\text{d}^3x\left\lbrace \frac{Na^5\rho}{2k^2}\left(\frac{\dot{\delta}_m^2}{N^2}+4\pi G_N\,\rho\,\delta_m^2\right)
+  \frac{Mb^5\tilde\rho}{2k^2}\left(\frac{\dot{\tilde\delta}_m^2}{M^2}+4\pi\,\tilde{G}_N\,\tilde\rho\,\tilde{\delta}_m^2\right)
\right\rbrace\,,
\end{equation}
where we have naturally defined $G_N = (8\pi\Mpl^2)^{-1}$ and $\tilde{G}_N = (8\pi\alpha^2\Mpl^2)^{-1}$ the Newton constants in both sectors.
So the no-ghost conditions ($\rho > 0$ and $\tilde\rho > 0$) and the equations of motion for the density contrasts
\begin{equation}\label{eq:eom_delta_SAbranch}
\frac{1}{N}\frac{\text{d}}{\text{d}t}\left(\frac{\dot{\delta}_m}{N}\right)+2H\, \frac{\dot{\delta}_m}{N}-4\pi G_N\rho\,\delta_m = 0
\qquad\text{and}\qquad
\frac{1}{M}\frac{\text{d}}{\text{d}t}\left(\frac{\dot{\tilde\delta}_m}{M}\right)+2L \,\frac{\dot{\tilde\delta}_m}{M}-4\pi \tilde{G}_N\tilde\rho\,\tilde{\delta}_m =0\,,
\end{equation}
are strictly the same as in GR.
It is thus clear that only two \emph{dof} propagates in the scalar sector, and that they can be identified as two matter \emph{dof}.
Each of the equations~\eqref{eq:eom_delta_SAbranch} reproduces exactly the phenomenology of a dust fluid in GR. 

When using the $k$-essence ansatz~\eqref{eq:Skess_def}, we also recover exactly two copies of the usual GR quadratic action, notably the no-ghost conditions and squared sound speeds are given exactly by~\eqref{eq:Skess_noghost_cs2}.
Therefore we can conclude that the phenomenology of the scalar sector of the self-accelerating branch is exactly given by two unrelated copies of GR.
Note that, contrary to the fate of the self-accelerating branch of HRBG, the self-accelerating branch of MTBG is perfectly healthy.

\subsubsection{Scalar sector in the normal branch}

In the scalar sector of the normal branch, the auxiliary field $\delta\ell$ enters linearly in the quadratic action, leading to the constraint
\begin{equation}
\zeta = \xi\,.
\end{equation}

As in the self-accelerating branch, we eliminate $j_0$ and $\tilde{j}_0$ in favor of the gauge invariant density contrasts~\eqref{eq:delta_m_j0}, and still have a remaining gauge freedom, that we use to fix $\chi = 0$.
One can then integrate out $j$ and $\tilde{j}$ \emph{via} the same relation as in the self-accelerating branch~\eqref{eq:deltaj_int_out}, so that the shift perturbations $\beta$ and $\theta$ still enter linearly and allow to express $\eta(v_m,\dot\zeta,\delta\lambda,\delta\bar\lambda)$ and $\omega(\tilde{v}_m,\dot\zeta,\delta\lambda,\delta\bar\lambda)$. At this point $\delta\lambda$ and $\delta\bar\lambda$ enter linearly and can be used to express $\zeta(\delta_m,\tilde{\delta}_m,v_m,\tilde{v}_m)$ and $\tau(\delta_m,\tilde{\delta}_m,v_m,\tilde{v}_m)$.
Finally, the fields $v_m$ and $\tilde{v}_m$ are not dynamical and can be integrated out to give a complicated action for $\delta_m$ and $\tilde{\delta}_m$, that are both dynamical.
Thus we can conclude that, also in the normal branch, the scalar sector of the theory propagates only two \emph{dof}, which can be identified as two matter \emph{dof}.

From a phenomenological point of view, let us take the sub-horizon (large $k$) limit.
At leading $k$ order, the action reads
\begin{equation}
\delta^{(2)}\mathcal{S}^S_\text{norm} = \int\!\text{d}t\text{d}^3x\left\lbrace \frac{Na^5\rho}{2k^2}\left(\frac{\dot{\delta}_m^2}{N^2}+4\pi \mathcal{G}_{11}\,\rho\,\delta_m^2\right)
+  \frac{Mb^5\tilde\rho}{2k^2}\left(\frac{\dot{\tilde\delta}_m^2}{M^2}+4\pi\,\mathcal{G}_{22}\,\tilde\rho\,\tilde{\delta}_m^2\right)
+ \frac{Na^3b^2\rho\,\tilde\rho}{k^2}\,\pi\,\mathcal{G}_{12}\,\delta_m\,\tilde{\delta}_m
\right\rbrace\,,
\end{equation}
as the kinetic coupling term $\propto \dot\delta_m\,\dot{\tilde\delta}_m$ and the friction term $\propto \dot\delta_m\,\tilde\delta_m- \delta_m\,\dot{\tilde\delta}_m$ appear at subleading $\mathcal{O}(k^{-4})$ order.
The no-ghost conditions ($\rho >0$ and $\tilde\rho>0$) are the same as in GR.
Nevertheless, the phenomenology of the scalar sector is different from the GR one, due to the non-standard mass terms.
Indeed, the components of the $\mathcal{G}$ matrix are given by
\begin{subequations}
\begin{align}
& \frac{\mathcal{G}_{11}}{G_N} = 1 + \frac{8m^2\left(c_1\mathcal{X}+c_2\right)\mathcal{X}^4H^2\left(1-r\right)
+4m^2\mathcal{X}^3H^2\Gamma_1
+m^4\left(\mathcal{X}^2+\alpha^{-2}\right)\left(r-2\right)\mathcal{X}^2\Gamma_1^2
-6m^2H^2\mathcal{X}^3\Gamma_1\,\Omega_m}{\left[4H^2\mathcal{X}-m^2\left(\mathcal{X}^2+\alpha^{-2}\right)\Gamma_1\right]^2}\,,\\
& \frac{\mathcal{G}_{12}}{G_N} =
\frac{16m^2H^2\mathcal{X}^3\left(c_1\mathcal{X}^2-c_3\right)\,r
-32m^2H^2\mathcal{X}^4\left(c_1\mathcal{X}+c_2\right)
+4m^4\mathcal{X}^2\left(\mathcal{X}^2+\alpha^{-2}\right)\Gamma_1^2
+24m^2H^2\mathcal{X}^3\Gamma_1\,\Omega_m}{\alpha^2\,\left[4H^2\mathcal{X}-m^2\left(\mathcal{X}^2+\alpha^{-2}\right)\Gamma_1\right]^2}\,,\\
& \frac{\mathcal{G}_{22}}{\tilde{G}_N} = 
1 + \frac{8m^2H^2\mathcal{X}\left(c_2\mathcal{X}+c_3\right)\,r
+4m^2H^2\mathcal{X}\left(c_1\mathcal{X}^2-c_3\right)
-m^4\left(\mathcal{X}^2+\alpha^{-2}\right)\,r\,\Gamma_1^2
-6m^2H^2\mathcal{X}\Gamma_1\,\Omega_m}{\alpha^2\,r\,\left[4H^2\mathcal{X}-m^2\left(\mathcal{X}^2+\alpha^{-2}\right)\Gamma_1\right]^2}\,,
\end{align}
\end{subequations}
with the $g$-sector matter density $\Omega_m \equiv \rho/(3\Mpl^2H^2)$, and the relations~\eqref{eq_cosmo_XR_evol} were used to eliminate $\tilde{\rho}$ in favor of the observable $\Omega_m$.

When using the $k$-essence ansatz~\eqref{eq:Skess_def}, we also recover, at the leading sub-horizon approximation, the usual no-ghost conditions and squared sound speeds~\eqref{eq:Skess_noghost_cs2}, and a non-standard mass matrix. It should be noticed that if the dynamics of the background leads to $4H^2\mathcal{X}-m^2\left(\mathcal{X}^2+\alpha^{-2}\right)\Gamma_1\to0$, then linear perturbation theory would break down, and in this case the system needs to be further studied. This possibility was present also in MTMG. 

Assuming in addition that the Hubble constant is notably larger than the mass of the massive graviton, as it is the case in the early Universe, one can expand the previous relations as
\begin{subequations}
\begin{align}
& \frac{\mathcal{G}_{11}}{G_N} = 1 + 
\frac{m^2\mathcal{X}}{2H^2}\left[\frac{3c_1\mathcal{X}^2+4c_2\mathcal{X}+c_3}{2}
-\left(c_1\mathcal{X}^2+c_2\mathcal{X}\right)\,r-\frac{3\Gamma_1}{4}\,\Omega_m\right]
+ \mathcal{O}\left(\frac{m^4}{H^4}\right)\,,\\
& \frac{\mathcal{G}_{12}}{G_N} =
\frac{m^2\mathcal{X}}{H^2\alpha^2}\left[\left(c_1\mathcal{X}^2-c_3\right)\,r-2\,\mathcal{X}\left(c_1\mathcal{X}+c_2\right)+\frac{3\Gamma_1}{2}\,\Omega_m\right]
+ \mathcal{O}\left(\frac{m^4}{H^4}\right)\,,\\
& \frac{\mathcal{G}_{22}}{\tilde{G}_N} = 1 +
\frac{m^2}{2H^2\alpha^2\,r\mathcal{X}}\left[\frac{c_1\mathcal{X}^2-c_3}{2}+\left(c_2\mathcal{X}+c_3\right)\,r-\frac{3\Gamma_1}{4}\,\Omega_m\right]
+ \mathcal{O}\left(\frac{m^4}{H^4}\right)\,.
\end{align}
\end{subequations}
Deviations from the usual phenomenology are thus strongly suppressed.
Nevertheless, if one seeks to mimic the late-time acceleration of the Universe by means of the effective dark fluid~\eqref{eq:eos_Normbranch}, it requires $m \sim H_0$, so the deviations can be large at the present epoch.

Considering only one dust component, minimally coupled in the $g$-sector, the action reads
\begin{equation}
\delta^{(2)}\mathcal{S}^S_\text{norm} = \int\!\text{d}t\text{d}^3x\frac{Na^5\rho}{2k^2}\left(\frac{\dot{\delta}_m^2}{N^2}+4\pi G_\text{eff}\,\rho\,\delta_m^2\right)\,.
\end{equation}
Once again, the no-ghost condition ($\rho >0$) is the same as in GR, and the dust experiences the effective Newton constant
\begin{equation}
\begin{aligned}
\frac{G_\text{eff}}{G_N} &\, =  1 
+ \frac{m^2\Gamma_1\mathcal{X}^2}{4H^2\mathcal{X}-m^2\left(\mathcal{X}^2+\alpha^{-2}\right)\Gamma_1}
+ \frac{24m^2H^4\mathcal{X}^4\left(c_1\mathcal{X}^2-c_3\right)\,\Omega_m}{\left[4H^2\mathcal{X}-m^2\left(\mathcal{X}^2+\alpha^{-2}\right)\Gamma_1\right]3}\\ 
& \, = 1 + 
\frac{m^2}{4H^2}\left[\Gamma_1\mathcal{X}+\frac{3\left(c_1\mathcal{X}^2-c_3\right)\mathcal{X}}{2}
\,\Omega_m\right]
+ \mathcal{O}\left(\frac{m^4}{H^4}\right)\,.
\end{aligned}
\end{equation}
Therefore we find that $G_{\rm eff}/G_N=1+\mathcal{O}(1)\times {m^2 \Mpl^2}/{\rho_m}$. The situation is similar to what happens in MTMG. The only gravity modes introduced by the theory are tensorial, no extra scalar degree of freedom is present, and $m$ is related to the mass of the massive graviton. Then whenever the mass-energy of the environment, $\rho_m$, becomes much larger than $m^2\Mpl^2$, we would naturally expect the graviton (sourced by matter fields) to be ultra-relativistic, that $G_{\rm eff}/G_N\to1$, and that dust matter fields in the dark sector coupled to the $f$-metric would decouple from the dark matter/baryonic components belonging to our physical sector coupled to the $g$-metric, at least, as long as linear perturbation theory is concerned. For a non-linear analysis of this effect in MTMG, please refer to \cite{MTMG_Nbody}.

\section{Conclusion and discussion}
\label{sec:concl}

Inspired by the principle of minimalism, we have constructed the minimal theory of bigravity (MTBG), a theory of two interacting spin-2 fields, that propagates only four degrees of freedom instead of the usual seven ones. This was achieved at the cost of a weak, i.e. cosmological-scale, breaking of the temporal component of the diffeomorphism invariance found in Hassan-Rosen bigravity (HRBG). 
Investigating the cosmology of MTBG, we found two branches of solutions for the background equations of motion.
Those branches are equivalent to the ones that are also present in HRBG, the only difference being that, in MTBG, they are both perfectly healthy and do not suffer from either strong-coupling issues or the Higuchi ghost.
In the first (self-accelerating) branch, the mass term produces two pure cosmological constants, one for each metric, and the phenomenology of the scalar and vector perturbations is the same as in GR. The tensor perturbations acquire an effective mass and oscillate, as usual in bigravity theories.
In  the background of the second (normal) branch, the mass term behaves as two dark fluids, with equations of state given by~\eqref{eq:eos_Normbranch}. While the vector sector is the same as in GR, the tensor perturbations acquire an effective mass and oscillate. More interestingly, the scalar perturbations have non-trivial phenomenology, with an effective Newton constant depending on the graviton mass, the background energy density and the scale of perturbations.

Let us also briefly discuss the possible gravitational Cherenkov effect in MTBG.
In fact, as it is clear from~\eqref{eq:tensor_pert}, gravitational perturbations propagate along light cones, so if $r$ is less than unity, the $f$-sector gravitational perturbations travel slower than the $g$-sector electromagnetic waves.
This would yield a gravitational Cherenkov effect, meaning that ultra-high-energy cosmic rays in the $g$-sector propagating at speeds greater than $c_f$ would loose their energy by emission of coherent $f$-gravitons.
As ultra-high-energy cosmic rays are observed on Earth, this mechanism could be used to put a lower bound on $r$~\cite{Grav_Cherenkov}.
However, such an effect has been investigated in detail in HRBG and the bounds on $(r,m^2)$ were found to be extremely weak~\cite{bigravity_Cherenkov}. 
An essential reason for this is that in HRBG, cosmic rays in the $g$-sector do not couple directly to the $f$-metric. 
Although we have not performed the detailed computation, we expect that a similar (if not the same) conclusion should hold in MTBG.  
Indeed the enforcement of the ADM vielbeins and the additional constraints that we use are not tensorial in nature, so it is expected that they would not significantly impact the computation of the Cherenkov effect.
Moreover the quadratic action of the tensor perturbations~\eqref{eq:tensor_pert} takes exactly the same form as the one in~\cite{bigravity_Cherenkov}, and the matter couplings are the same in both works.
Therefore we expect that the gravitational Cherenkov radiation would not put stringent bounds on the parameter space of MTBG.

It would be interesting to investigate the normal branch more deeply, to see how well this modified gravitational strength can accommodate the most recent cosmological data.
As for the self-accelerating one, it provides the simplest testing ground for graviton oscillations~\cite{DeFelice:2013nba} and massive graviton dark matter~\cite{Aoki:2016zgp} as well as the enhancement mechanism of stochastic gravitational waves~\cite{Fujita:2018ehq}.
Indeed it perfectly mimics the $\Lambda$-CDM model at the background and scalar/vector perturbative levels, but has a non-canonical tensor sector.
On the other hand, it would also be important to investigate its strong field regime, e.g.~by constructing solutions of compact objects and seek phenomenological discrepancies with respect to GR.
Those investigations are naturally left for future work.

Concerning the matter sector, we have made the simplest choice in this work : two minimally coupled perfect fluids or dust components. 
However our construction could accommodate more sophisticated matter couplings, as long as our particular constraint algebra is preserved.
One could think, for example, of a matter sector coupled to a composite metric, with or without the specific structure studied in~\cite{deRham:2014naa,Gumrukcuoglu:2015nua,DeFelice:2015yha}.

\acknowledgments

MO would like to express his gratitude to R.\ Durrer for her
hospitality during an initial phase of the project.  FL would like to
thank L.\ Pinol for enlightening discussions about technical details
on cosmological computations. The work of ADF was supported by Japan
Society for the Promotion of Science Grants-in-Aid for Scientific
Research No.~20K03969. The work of SM was supported in part by Japan
Society for the Promotion of Science Grants-in-Aid for Scientific
Research No.~17H02890, No.~17H06359, and by World Premier
International Research Center Initiative, MEXT, Japan.


\appendix


\section{Hamiltonian analysis of the precursor theory}
\label{app:Ham_pre}

In this appendix we show the detailed computation of the constraint algebra of the precursor Hamiltonian~\eqref{eq:precursor_H}.

\subsection{Internal algebra} 

The diagonal blocks of the internal algebra of the $g$-sector reads
\begin{subequations}
\begin{align}
& \left\lbrace\mathcal{R}_0\left[\zeta\right],\mathcal{R}_0\left[\eta\right]\right\rbrace =
\int\!\dd^3y\,\mathcal{R}^i\left(\eta\mathcal{D}_i\zeta-\zeta\mathcal{D}_i\eta\right) \approx 0\,,\\
& \left\lbrace\mathcal{R}_i\left[\xi^i\right],\mathcal{R}_j\left[\chi^j\right]\right\rbrace =
\int\!\dd^3y\,\mathcal{R}_j\left(\chi^i\mathcal{D}_i\xi^j-\xi^i\mathcal{D}_i\chi^j\right)  \approx 0\,.
\end{align}
\end{subequations}

The off-diagonal one is
\begin{equation}
\begin{aligned}
\left\lbrace \mathcal{R}_0\left[\zeta\right], \mathcal{R}_i\left[\xi^i\right]\right\rbrace & = \int\!\dd^3y \mathcal{R}_0^\text{GR}\,\xi ^i\mathcal{D}_i\zeta 
+ \frac{m^2\Mpl^2}{2}\int\!\dd^3y\sqrt\gamma\,\mathcal{H}_0\,\zeta\mathcal{D}_i\xi^i
+ m^2\Mpl^2\int\!\dd^3y\sqrt\gamma\,\frac{\partial\mathcal{H}_0}{\partial\gamma_{ij}}\,\gamma_{ik}\,\zeta\mathcal{D}_j\xi^k\\
& =
\int\!\dd^3y \mathcal{R}_0\,\xi ^i\mathcal{D}_i\zeta 
- \frac{m^2\Mpl^2}{2}\int\!\dd^3y\sqrt\gamma\,\zeta\xi^i\,\mathcal{D}_i\mathcal{H}_0
+ m^2\Mpl^2\int\!\dd^3y\sqrt\gamma\,\frac{\partial\mathcal{H}_0}{\partial\gamma_{ij}}\,\gamma_{ik}\,\zeta\mathcal{D}_j\xi^k\,.
\end{aligned}
\end{equation}
After a straightforward derivation, and independently of the commutativity properties of $\gamma_{ij}$ and $\phi_{ij}$, it follows that
\begin{equation}
\frac{\partial e_n(\mathfrak{K})}{\partial \gamma_{ij}} = -\frac{1}{2}\gamma^{q(i}\mathfrak{K}^{j)}_{\ p}\,U^p_{(n)\,q}\,,
\qquad \text{thus} \qquad
\frac{\partial \mathcal{H}_0}{\partial \gamma_{ij}} 
= -\frac{1}{2} \sum_{n=0}^3 c_{4-n} \gamma^{q(i}\mathfrak{K}^{j)}_{\ p}\,U^p_{(n)\,q}\,,
\end{equation}
and that
\begin{equation}
\frac{\partial\left(\sqrt\gamma\mathcal{H}_0\right)}{\partial\gamma_{ij}}
 = \frac{\sqrt\gamma}{2}\sum_{n=0}^3c_{4-n} \left(e_n\,\gamma^{ij} - \gamma^{q(i}\mathfrak{K}^{j)}_{\ p}\,U^p_{(n)\,q}\right)
= \frac{\sqrt\gamma}{2}\sum_{n=0}^3c_{4-n} \gamma^{q(i}U^{j)}_{(n+1)\,q}\,.
\end{equation}
As $U_{(4)} = 0$, the constraint finally reads
\begin{equation}
\left\lbrace \mathcal{R}_0\left[\zeta\right], \mathcal{R}_i\left[\xi^i\right]\right\rbrace \approx
- \frac{m^2\Mpl^2}{2}\int\!\dd^3y\sqrt\gamma\,\zeta\,\mathcal{D}_i\left(\mathcal{H}_0\xi^i\right)
+ \frac{m^2\Mpl^2}{4}\int\!\dd^3y\sqrt\gamma\,\sum_{n=1}^3c_{5-n} \left(U^i_{(n)\,j} + \gamma^{iq}\gamma_{jp}U^p_{(n)\,q}\right)\zeta\,\mathcal{D}_i\xi^j\,.
\end{equation}

Similarly, in the $f$-sector we have
\begin{subequations}
\begin{align}
& \left\lbrace\tilde{\mathcal{R}}_0\left[\zeta\right],\tilde{\mathcal{R}}_0\left[\eta\right]\right\rbrace \approx
\left\lbrace\tilde{\mathcal{R}}_i\left[\xi^i\right],\tilde{\mathcal{R}}_j\left[\chi^j\right]\right\rbrace\approx 0\,,\\
&  
\left\lbrace\tilde{\mathcal{R}}_0\left[\zeta\right],\tilde{\mathcal{R}}_i\left[\xi^i\right]\right\rbrace \approx 
- \frac{m^2\Mpl^2}{2}\int\!\dd^3y\sqrt\phi\,\zeta\,\mathfrak{D}_i\left(\tilde{\mathcal{H}}_0\xi^i\right)
+\frac{m^2\Mpl^2}{4}\int\!\dd^3y\sqrt\phi
\sum_{n=1}^3c_{n-1}\left( V_{(n)\, j}^i +\phi^{iq}\phi_{jp} \, V_{(n)\, q}^p
\right)\zeta\mathfrak{D}_i\xi^j\,.
\end{align}
\end{subequations}

\subsection{Cross-algebra} 

The ``cross-algebra'' can be easily computed after a first step: following the previous argument together with $\sqrt\phi \, e_{n}(\mathcal{K})=\sqrt\gamma \, e_{3-n} (\mathfrak{K}) $, it follows that
\begin{equation}
\sqrt\phi \, \frac{\partial \tilde{\mathcal{H}}_0}{\partial \gamma_{ij}} 
=  \frac{\partial}{\partial\gamma_{ij}}\left[\sqrt\phi \,\tilde{\mathcal{H}}_0\right]
=  \frac{\partial}{\partial\gamma_{ij}}\left[\sqrt\gamma\sum_{n=0}^3c_{3-n}e_n(\mathfrak{K})\right]
= \frac{\sqrt\gamma}{2}\sum_{n=1}^3c_{4-n}\gamma^{p(i}U_{(n)\, p}^{j)}\,,
\end{equation} 
and that
\begin{equation}
\sqrt\gamma \, \frac{\partial \mathcal{H}_0}{\partial \phi_{ij}} = \frac{\sqrt\phi}{2}\sum_{n=1}^3c_n\,\phi^{p(i}V_{(n)\, p}^{j)}\,.
\end{equation}
Therefore, one has
\begin{subequations}
\begin{align}
& \left\lbrace\mathcal{R}_0\left[\zeta\right],\tilde{\mathcal{R}}_0\left[\eta\right]\right\rbrace 
= m^2\int\!\dd^3y\,
\sum_{n=1}^3\left[\frac{c_n}{\alpha^2}\left(\sigma^i_j-\frac{\sigma}{2}\delta^i_j\right)V^j_{(n)\,i}-c_{4-n}\left(\pi^i_j-\frac{\pi}{2}\delta^i_j\right)U^j_{(n)\,i}\right]\zeta\eta
\,,\\
& \left\lbrace\mathcal{R}_0\left[\zeta\right],\tilde{\mathcal{R}}_i\left[\xi^i\right]\right\rbrace = \frac{m^2\Mpl^2}{4}\int\!\dd^3y\sqrt\phi\,
\sum_{n=1}^3c_n\left( V_{(n)\, j}^i +\phi^{iq}\phi_{jp} \, V_{(n)\, q}^p\right)\zeta\mathfrak{D}_i\xi^j\,,\\
& \left\lbrace\mathcal{R}_i\left[\xi^i\right],\tilde{\mathcal{R}}_0\left[\zeta\right]\right\rbrace = 
\frac{m^2\Mpl^2}{4}\int\!\dd^3y\sqrt\gamma\,\zeta
\sum_{n=1}^3c_{4-n}\left( U_{(n)\, j}^i +\gamma^{iq}\gamma_{jp} \, U_{(n)\, q}^p
\right)\zeta\mathcal{D}_i\xi^j\,,\\
& \left\lbrace\mathcal{R}_i\left[\xi^i\right],\tilde{\mathcal{R}}_j\left[\chi^j\right]\right\rbrace =0\,.
\end{align}
\end{subequations}


\section{Residual spatial diffeomorphism in precursor theory and MTBG}
\label{app:Ham_MTBG}

In this appendix, we prove that the combinations~\eqref{eq:def_Ri_3D} 
\begin{equation}
\mathrm{R}_i \equiv \mathcal{R}_i+\tilde{\mathcal{R}}_i\,,
\end{equation}
are first-class constraints in the precursor theory and MTBG, by computing their Poisson brackets with all independent constraints. Obviously, these three first-class constraints are generators of the spatial diffeomorphism under which both the precursor theory and MTBG are invariant. 
In either theory, as none of the momenta conjugate to the Lagrange multipliers $\{N,N^i,M,M^i,\lambda,\bar\lambda,\lambda^i\}$ enter the Hamiltonian~\eqref{eq:precursor_H} or \eqref{eq:MTBG_H}, we can naturally restrict the Poisson brackets to operate on $\{\gamma_{ij},\phi_{ij}\}$ and their conjugate momenta.
In addition, as $\mathrm{R}_i$ are invariant under the exchange of $g$- and $f$-sectors, we will only explicitly present the derivation of its Poisson brackets with $g$-sector quantities, the computation of the other Poisson brackets being very similar.

Since $\{\mathcal{R}_0,\mathcal{R}_i,\tilde{\mathcal{R}}_0,\tilde{\mathcal{R}}_i\}$ take the same expressions in the precursor theory as in MTBG, the following analysis in the first two subsections of this appendix provides a proof that $\mathrm{R}_i$ are first-class constraints of the precursor theory. In MTBG, since there are additional constraints, one needs to compute more Poisson brackets than in the precursor theory. This is what we shall do in the other subsections.

\subsection{Poisson bracket with $\mathcal{R}_0$ and $\tilde{\mathcal{R}_0}$}

Using the results of App.~\ref{app:Ham_pre}, let us decompose
\begin{equation}
\begin{aligned}
\left\lbrace \mathcal{R}_0\left[\zeta\right], \mathrm{R}_i\left[\xi^i\right]\right\rbrace
& = \left\lbrace \mathcal{R}_0\left[\zeta\right], \mathcal{R}_i\left[\xi^i\right]\right\rbrace+\left\lbrace \mathcal{R}_0\left[\zeta\right], \tilde{\mathcal{R}}_i\left[\xi^i\right]\right\rbrace\\
&  =
\int\!\dd^3y \mathcal{R}_0\,\xi ^i\mathcal{D}_i\zeta 
- \frac{m^2\Mpl^2}{2}\int\!\dd^3y\sqrt\gamma\,\zeta\,\mathcal{D}_i\left(\mathcal{H}_0\xi^i\right)\\
& \quad
+ \frac{m^2\Mpl^2}{4}\int\!\dd^3y\zeta\sum_{n=1}^3\left[\sqrt\gamma\,c_{5-n} \left(U^i_{(n)\,j} + \gamma^{iq}\gamma_{jp}U^p_{(n)\,q}\right)\mathcal{D}_i\xi^j
+\sqrt\phi\,
c_n\left( V_{(n)\, j}^i +\phi^{iq}\phi_{jp} \, V_{(n)\, q}^p\right)\mathfrak{D}_i\xi^j\right]\\
& = 
\int\!\dd^3y \mathcal{R}_0\,\xi ^i\mathcal{D}_i\zeta
+ \frac{m^2\Mpl^2c_1}{2}\int\!\dd^3y\sqrt\phi\,\zeta\left(\mathfrak{D}_i\xi^i-\mathcal{D}_i\xi^i-\frac{1}{2}\,\xi^i\phi^{pq}\mathcal{D}_i\phi_{pq}\right)\\
& \quad
+ \frac{m^2\Mpl^2c_2}{4}\int\!\dd^3y\sqrt\gamma\,\zeta\left[\gamma^{pq}\xi^i\mathcal{D}_i\phi_{pq} + 2 \gamma^{pr}\phi_{rq}\left(\mathcal{D}_p\xi^q-\mathfrak{D}_p\xi^q\right)\right.\\
& \hspace{5cm}\left.
-2\mathfrak{K}^q_{\, q}\xi^i\mathcal{D}_i\mathfrak{K}^p_{\, p} -\mathfrak{K}^r_{\, r} \mathfrak{K}^p_{\ q}\left(\mathcal{D}_p\xi^q-\mathfrak{D}_p\xi^q+\gamma_{ip}\gamma^{jq}\mathcal{D}_j\xi^i-\phi_{ip}\phi^{jq}\mathfrak{D}_j\xi^i\right)\right]\\
& \quad
- \frac{m^2\Mpl^2c_3}{4}\int\!\dd^3y\sqrt\gamma\,\zeta\left[2\xi^i\mathcal{D}_i\mathfrak{K}^p_{\, p} + \mathfrak{K}^p_{\ q}\left(\mathcal{D}_p\xi^q-\mathfrak{D}_p\xi^q+\gamma_{ip}\gamma^{jq}\mathcal{D}_j\xi^i-\phi_{ip}\phi^{jq}\mathfrak{D}_j\xi^i\right)\right]\,,
\end{aligned}
\end{equation}
where we have used the relation $\sqrt\phi\, V^p_{(n)\, q} = \sqrt\gamma\, U^p_{(4-n)\, r}\mathfrak{K}^r_{\ q}$.
Using a local inertial frame, we can write
\begin{equation}
\mathcal{D}_p\xi^q-\mathfrak{D}_p\xi^q = \partial_p\xi^q - \mathfrak{D}_p\xi^q = -\frac{1}{2}\,\phi^{qr}\left(\mathcal{D}_p\phi_{rs}+\mathcal{D}_s\phi_{pr}-\mathcal{D}_r\phi_{ps}\right)\xi^s\,,
\end{equation}
so that the term proportional to $c_1$ and the first part of the term proportional to $c_2$ vanish. Next, developing $\mathcal{D}_i(\mathfrak{K}^p_{\ r}\mathfrak{K}^r_{\ q}) = \gamma^{ps}\mathcal{D}_i\phi_{sq}$, it follows that $2\mathcal{D}_i\mathfrak{K}^p_{\ p} = \mathcal{K}^{a}_{\ b}\gamma^{bc}\mathcal{D}_i\phi_{ac}= \mathfrak{K}^{a}_{\ b}\phi^{bc}\mathcal{D}_i\phi_{ac}$, thus the remaining part is proportional to
\begin{equation}
\begin{aligned}
& 2\xi^i\mathcal{D}_i\mathfrak{K}^p_{\, p} + \mathfrak{K}^p_{\ q}\left(\mathcal{D}_p\xi^q-\mathfrak{D}_p\xi^q+\gamma_{ip}\gamma^{jq}\mathcal{D}_j\xi^i-\phi_{ip}\phi^{jq}\mathfrak{D}_j\xi^i\right)\\
& \quad = \xi^i\mathfrak{K}^p_{\, r}\phi^{rq}\mathcal{D}_i\phi_{pq} -\frac{\mathfrak{K}^p_{\ r}\phi^{rq}}{2}\left(\mathcal{D}_p\phi_{rs}+\mathcal{D}_s\phi_{rp}-\mathcal{D}_r\phi_{ps}\right)\xi^s\\
& \quad \quad + \mathfrak{K}^p_{\ q}\left[\left(\gamma_{ip}\gamma^{jq}-\phi_{ip}\phi^{jq}\right)\mathcal{D}_j\xi^i- \frac{\phi^{jq}}{2}\left(\mathcal{D}_j\phi_{ps}+\mathcal{D}_s\phi_{pj}-\mathcal{D}_p\phi_{js}\right)\xi^s\right]\\
& \quad = \left(\gamma_{ip}\mathfrak{K}^p_{\ q}\gamma^{jq}-\phi_{ip}\mathfrak{K}^p_{\ q}\phi^{jq}\right)\mathcal{D}_j\xi^i\\
& \quad =\left(\gamma_{ip}\mathfrak{K}^p_{\ q}-\phi_{ip}\mathcal{K}^p_{\ q}\right)\gamma^{jq}\mathcal{D}_j\xi^i\\
& \quad = 0\,.
\end{aligned}
\end{equation}

So finally
\begin{equation}
\left\lbrace \mathcal{R}_0\left[\zeta\right], \mathrm{R}_i\left[\xi^i\right]\right\rbrace
 =
\int\!\dd^3y \mathcal{R}_0\,\xi ^i\mathcal{D}_i\zeta 
\approx 0\,,
\end{equation}
and a similar reasoning leads to
\begin{equation}
\left\lbrace \tilde{\mathcal{R}}_0\left[\zeta\right], \mathrm{R}_i\left[\xi^i\right]\right\rbrace
 =
\int\!\dd^3y \tilde{\mathcal{R}}_0\,\xi ^i\mathcal{D}_i\zeta 
\approx 0\,.
\end{equation}

\subsection{Poisson bracket with $\mathcal{R}_i$ and $\mathrm{R}_i$}

Let us decompose
\begin{equation}
\left\lbrace \mathcal{R}_j\left[\chi^j\right], \mathrm{R}_i\left[\xi^i\right]\right\rbrace 
=\left\lbrace \mathcal{R}_j\left[\chi^j\right], \mathcal{R}_i\left[\xi^i\right]\right\rbrace 
+ \left\lbrace \mathcal{R}_j\left[\chi^j\right], \tilde{\mathcal{R}}_i\left[\xi^i\right]\right\rbrace\,.
\end{equation}
It follows that
\begin{equation}
\left\lbrace \mathcal{R}_j\left[\chi^j\right], \mathcal{R}_i\left[\xi^i\right]\right\rbrace = \int\!\text{d}^3x\,\mathcal{R}_b\left(\xi^a\mathcal{D}_a\chi^b-\chi^a\mathcal{D}_a\xi^b\right)\,,
\qquad\text{and}\qquad
\left\lbrace \mathcal{R}_j\left[\chi^j\right], \tilde{\mathcal{R}}_i\left[\xi^i\right]\right\rbrace = 0\,,
\end{equation}
so that
\begin{equation}\label{eq:PB_Ri_Ri}
\left\lbrace \mathcal{R}_j\left[\chi^j\right], \mathrm{R}_i\left[\xi^i\right]\right\rbrace 
=
\int\!\text{d}^3x\,\mathcal{R}_b\left(\xi^a\partial_a\chi^b-\chi^a\partial_a\xi^b\right) \approx 0\,.
\end{equation}

Using the fact that $\xi^a\mathcal{D}_a\chi^b-\chi^a\mathcal{D}_a\xi^b = \xi^a\partial_a\chi^b-\chi^a\partial_a\xi^b =\xi^a\mathfrak{D}_a\chi^b-\chi^a\mathfrak{D}_a\xi^b= \left[\vec{\chi},\vec{\xi}\right]^b$, where the vector commutator $\left[\;,\;\right]$ does not depend on the used metric for vectors, and the natural generalization of~\eqref{eq:PB_Ri_Ri} to the $f$-sector, it follows that
\begin{equation}
\left\lbrace \mathrm{R}_j\left[\chi^j\right], \mathrm{R}_i\left[\xi^i\right]\right\rbrace 
=
\int\!\text{d}^3x\,\mathrm{R}_b\left(\xi^a\partial_a\chi^b-\chi^a\partial_a\xi^b\right) \approx 0\,.
\end{equation}

\subsection{Poisson bracket with $\mathcal{C}_0 - \tilde{\mathcal{C}}_0$}

Let us recall that $\mathcal{C}_0$ was defined in~\eqref{eq:def_C0_Ci} by
\begin{equation}
\mathcal{C}_0\left[\zeta\right] \equiv \left\lbrace\mathcal{R}_0^\text{GR}\left[\zeta\right],\tilde\Theta\right\rbrace = \left\lbrace\mathcal{R}_0\left[\zeta\right],\tilde\Theta\right\rbrace\,,
\qquad\text{where}\qquad
\tilde\Theta \equiv -\frac{m^2\Mpl^2}{2}\int\!\text{d}^3x\,\sqrt\phi \tilde{\mathcal{H}}_0\,.
\end{equation}
Let us first compute
\begin{equation}
\begin{aligned}
\left\lbrace\tilde\Theta, \mathrm{R}_i\left[\xi^i\right]\right\rbrace
& = m^2\Mpl^2\int\!\dd^3y\left[\frac{\partial(\sqrt\phi\tilde{\mathcal{H}}_0)}{\partial\gamma_{ij}}\gamma_{ik}\mathcal{D}_j\xi^k+\frac{\partial(\sqrt\phi\tilde{\mathcal{H}}_0)}{\partial\phi_{ij}}\phi_{ik}\mathfrak{D}_j\xi^k\right]\\
& = \frac{m^2\Mpl^2}{2}\int\!\dd^3y\sqrt\gamma\sum_{n=1}^4c_{4-n}\left[\gamma^{p(i}U^{j)}_{(n)\ p}\gamma_{ik}\mathcal{D}_j\xi^k+\phi^{p(i}U^{j)}_{(n-1)\ q}\mathfrak{K}^q_{\ p}\phi_{ik}\mathfrak{D}_j\xi^k\right]\\
& = \frac{m^2\Mpl^2}{2}\int\!\dd^3y\sqrt\gamma\sum_{n=2}^3c_{4-n}\left[e_n\left(\mathfrak{K}\right)\mathcal{D}_i\xi^i+ U^{p}_{(n-1)\, r}\mathfrak{K}^r_{\ q}\left(\phi^{sq}\phi_{k(s}\mathfrak{D}_{p)}\xi^k-\gamma^{sq}\gamma_{k(s}\mathcal{D}_{p)}\xi^k\right)\right]\\
& = 0 \,,
\end{aligned}
\end{equation}
where we have used the fact that $U^p_{(n)\, q} = e_n\left(\mathfrak{K}\right)\,\delta^p_q-U^p_{(n-1)\, r}\mathfrak{K}^r_{\ q}$  and the last equality follows from the same type of reasoning that was used when computing $\left\lbrace \mathcal{R}_0\left[\zeta\right], \mathrm{R}_i\left[\xi^i\right]\right\rbrace$.
Let us note that this Poisson bracket vanishes \emph{off-shell}.

Next, by using the Jacobi identity, we can rewrite
\begin{equation}
\begin{aligned}
\left\lbrace\mathcal{C}_0\left[\zeta\right], \mathrm{R}_i\left[\xi^i\right]\right\rbrace 
& = \left\lbrace\left\lbrace \mathcal{R}_0\left[\zeta\right],\tilde\Theta\right\rbrace, \mathrm{R}_i\left[\xi^i\right]\right\rbrace \\
& = -\left\lbrace\left\lbrace\mathrm{R}_i\left[\xi^i\right], \mathcal{R}_0\left[\zeta\right]\right\rbrace,\tilde\Theta \right\rbrace
-\left\lbrace\left\lbrace\tilde\Theta,\mathrm{R}_i\left[\xi^i\right]\right\rbrace, \mathcal{R}_0\left[\zeta\right] \right\rbrace\\
& = \left\lbrace \int\!\text{d}^3x\,\mathcal{R}_0\,\xi^a\mathcal{D}_a\zeta,\tilde\Theta\right\rbrace\\
& = \int\!\text{d}^3x\,\mathcal{C}_0\,\xi^a\mathcal{D}_a\zeta\\
& = \mathcal{C}_0\left[\xi^a\mathcal{D}_a\zeta\right]\,,
\end{aligned}
\end{equation}
where we have used the fact that $\tilde\Theta$ does not depend on $\pi^{ij}$ in the last-but-one line. A similar path leads to
\begin{equation}
\left\lbrace\tilde{\mathcal{C}}_0\left[\zeta\right], \mathrm{R}_i\left[\xi^i\right]\right\rbrace =
\tilde{\mathcal{C}}_0\left[\xi^a\mathfrak{D}_a\zeta\right]\,,
\end{equation}
Using the fact that $\zeta$ is a scalar, so $\mathcal{D}_i\zeta = \mathfrak{D}_i\zeta = \partial_i\zeta$, we have
\begin{equation}
\left\lbrace\left(\mathcal{C}_0-\tilde{\mathcal{C}}_0\right)\left[\zeta\right], \mathrm{R}_i\left[\xi^i\right]\right\rbrace 
 = \left\lbrace\mathcal{C}_0\left[\zeta\right], \mathrm{R}_i\left[\xi^i\right]\right\rbrace -\left\lbrace\tilde{\mathcal{C}}_0\left[\zeta\right], \mathrm{R}_i\left[\xi^i\right]\right\rbrace 
 = \left(\mathcal{C}_0-\tilde{\mathcal{C}}_0\right)\left[\zeta\right]\approx 0\,.
\end{equation}

\subsection{Poisson bracket with $\mathcal{C}_i-\beta\,\tilde{\mathcal{C}}_i$}

Let us recall that $\mathcal{C}_i$ was defined in~\eqref{eq:def_C0_Ci} by
\begin{equation}
\mathcal{C}_i\left[\xi^i\right] = \left\lbrace \mathcal{R}_i\left[\xi^i\right],\tilde\Theta\right\rbrace\,,
\end{equation}
where $\tilde{\Theta}$ has been defined just above.
Using the Jacobi identity, and the fact that $\{\tilde\Theta, \mathrm{R}_i\left[\xi^i\right]\} = 0$, we can rewrite
\begin{equation}
\begin{aligned}
\left\lbrace\mathcal{C}_j\left[\chi^j\right], \mathrm{R}_i\left[\xi^i\right]\right\rbrace 
& = \left\lbrace\left\lbrace \mathcal{R}_j\left[\chi^j\right],\tilde\Theta\right\rbrace, \mathrm{R}_i\left[\xi^i\right]\right\rbrace \\
& = -\left\lbrace\left\lbrace\mathrm{R}_i\left[\xi^i\right], \mathcal{R}_j\left[\chi^j\right]\right\rbrace,\tilde\Theta \right\rbrace
-\left\lbrace\left\lbrace\tilde\Theta,\mathrm{R}_i\left[\xi^i\right]\right\rbrace, \mathcal{R}_j\left[\chi^j\right] \right\rbrace\\
& = \left\lbrace \int\!\text{d}^3x\,\mathcal{R}_b\left(\xi^a\partial_a\chi^b-\chi^a\partial_a\xi^b\right),\tilde\Theta\right\rbrace\\
& = \int\!\text{d}^3x\,\mathcal{C}_b\left(\xi^a\partial_a\chi^b-\chi^a\partial_a\xi^b\right)\\
& = \mathcal{C}_b\left[\xi^a\partial_a\chi^b-\chi^a\partial_a\xi^b\right]\,.
\end{aligned}
\end{equation}
A similar path leads to
\begin{equation}
\left\lbrace\tilde{\mathcal{C}}_j\left[\chi^j\right], \mathrm{R}_i\left[\xi^i\right]\right\rbrace =
\tilde{\mathcal{C}}_b\left[\xi^a\partial_a\chi^b-\chi^a\partial_a\xi^b\right]\,,
\end{equation}
so that, as $\beta$ is a constant,
\begin{equation}
\left\lbrace\left(\mathcal{C}_j-\beta\,\tilde{\mathcal{C}}_j\right)\left[\chi^j\right], \mathrm{R}_i\left[\xi^i\right]\right\rbrace 
 = \left\lbrace\mathcal{C}_j\left[\chi^j\right], \mathrm{R}_i\left[\xi^i\right]\right\rbrace -\beta\,\left\lbrace\tilde{\mathcal{C}}_j\left[\chi^j\right], \mathrm{R}_i\left[\xi^i\right]\right\rbrace 
 = \left(\mathcal{C}_j-\beta\,\tilde{\mathcal{C}}_j\right)\left[\xi^a\partial_a\chi^j-\chi^a\partial_a\xi^j\right]\approx 0\,.
\end{equation}

\subsection{Poisson bracket with $\sqrt\gamma \gamma^{ij} \mathcal{D}_{ij}(\mathcal{C}_0/\sqrt\gamma)+\sqrt\phi \phi^{ij} \mathfrak{D}_{ij}(\tilde{\mathcal{C}}_0/\sqrt\phi)$}

After some manipulation, it comes
\begin{equation}
\left\lbrace\sqrt\gamma \gamma^{pq} \mathcal{D}_{pq}\left(\frac{\mathcal{C}_0}{\sqrt\gamma}\right)\left[\zeta\right], \mathrm{R}_i\left[\xi^i\right]\right\rbrace  =
\left\lbrace \int\!\dd^3y \, \mathcal{C}_0\, \gamma^{pq} \mathcal{D}_{pq}\zeta , \mathrm{R}_i\left[\xi^i\right]\right\rbrace  =
\int\!\dd^3y\sqrt\gamma\,\gamma^{pq}\mathcal{D}_{pq}\left(\frac{\mathcal{C}_0}{\sqrt\gamma}\right)\,\xi^i\partial_i\zeta\,,
\end{equation}
and similarly in the $g$-sector, so that
\begin{equation}
\left\lbrace\left(\sqrt\gamma \gamma^{pq} \mathcal{D}_{pq}\left(\frac{\mathcal{C}_0}{\sqrt\gamma}\right)+\sqrt\phi \phi^{pq} \mathfrak{D}_{pq}\left(\frac{\tilde{\mathcal{C}}_0}{\sqrt\phi}\right)\right)\left[\zeta\right], \mathrm{R}_i\left[\xi^i\right]\right\rbrace  =
\left(\sqrt\gamma \gamma^{pq} \mathcal{D}_{pq}\left(\frac{\mathcal{C}_0}{\sqrt\gamma}\right)+\sqrt\phi \phi^{pq} \mathfrak{D}_{pq}\left(\frac{\tilde{\mathcal{C}}_0}{\sqrt\phi}\right)\right)\left[\xi^i\partial_i\zeta\right] \approx 0\,.
\end{equation}

\subsection{Summary}

For a generic constraint $\mathcal{C}$ entering the Hamiltonian~\eqref{eq:precursor_H} or \eqref{eq:MTBG_H}, it follows that
\begin{equation}
\left\lbrace\mathcal{C},\mathrm{R}_i\right\rbrace \propto \mathcal{C} \approx 0\,,
\end{equation}
so that $\mathrm{R}_i$ are indeed first-class constraints, removing $2\times 3 = 6$ phase space \emph{dof}.


\section{MTBG in the vielbein language}\label{app:Vielbein}

Let us introduce two sets of spatial vielbeins $\{e^I_{\ j},E^I_{\ j}\}$, one for each metric, so that
\begin{equation}
\gamma_{ij} = \delta_{IJ}\,e^I_{\ i}e^J_{\ j}\,,
\qquad \text{and} \qquad
\phi_{ij} = \delta_{IJ}\,E^I_{\ i}E^J_{\ j}\,.
\end{equation}
It is also convenient to introduce the dual basis $\{e_I^{\ j},E_I^{\ j}\}$ such that
\begin{equation}
e^I_{\ k}\,e_J^{\ k} = \delta^I_J\,,
\qquad
e^K_{\ j}\,e_K^{\ i} = \delta^i_j\,,
\qquad
E^I_{\ k}\,E_J^{\ k} = \delta^I_J\,,
\qquad\text{and}\qquad
E^K_{\ j}\,E_K^{\ i} = \delta^i_j\,,
\end{equation}
and to construct the mixed products
\begin{equation}
X_I^{\ J} \equiv e_I^{\ k} E^J_{\ k}\,,
\qquad\text{and}\qquad
Y_I^{\ J} \equiv E_I^{\ k} e^J_{\ k}\,,
\end{equation}
corresponding to the square-root matrices~\eqref{eq:gothic_K_def} in the metric formulation.

\subsection{Precursor theory}

In the vielbein language, the precursor action~\eqref{eq:precursor_action} is written as
\begin{equation}
\begin{aligned}
\mathcal{S}_\text{pre} = 
\frac{\Mpl^2}{2}\int\!\dd^4x &\left\lbrace 
\sqrt{-g} \mathcal{R}\left[g\right]
+ \alpha^2 \sqrt{-f} \mathcal{R}\left[f\right] 
- m^2\left[
 c_0\sqrt\phi M + c_1\sqrt\phi\left(N + M Y_I^{\ I}\right)\right.\right. \\
& \left.\left.
+ c_2\sqrt\phi\left(NY_I^{\ I} + M \frac{Y_I^{\ I}Y_J^{\ J}-Y_I^{\ J}Y_J^{\ I}}{2}\right)
+ c_3\sqrt\gamma\left(M + N X_I^I\right) + c_4\sqrt\gamma N \right]\right\rbrace \,,
\end{aligned}
\end{equation}
and so the primary Hamiltonian becomes
\begin{equation}
H_\text{pre}^{(1)} = -\int\!\dd^3x\left\lbrace N\mathcal{R}_g+N^i\mathcal{R}_i+M\mathcal{R}_f +M^i\tilde{\mathcal{R}}_i + \alpha_{MN}\,P^{[MN]}+\beta_{MN}\,Q^{[MN]}\right\rbrace\,.
\end{equation}
where 
\begin{subequations}
\begin{align}
& \mathcal{R}_g = \mathcal{R}_0^\text{GR}- \frac{\Mpl^2m^2}{2}\mathcal{H}_g\,,\qquad
 \mathcal{H}_g = \sqrt\phi\left(c_1 + c_2\,Y_I^{\ I}\right) + \sqrt\gamma\left(c_3 X_I^{\ I} + c_4\right)\,,\\
& \mathcal{R}_f =  \tilde{\mathcal{R}}_0^\text{GR}- \frac{\Mpl^2m^2}{2}\mathcal{H}_f\,,\qquad
\mathcal{H}_f = \sqrt\phi\left(c_0 + c_1\,Y_I^{\ I}\right) + \sqrt\gamma\left(c_2 X_I^{\ I} + c_3\right)\,,\\
& P^{[MN]} = e^M_{\ i}\,\Pi_K^{\ i}\,\delta^{KN}- e^N_{\ i}\,\Pi_K^{\ i}\,\delta^{KM}\,, \qquad
 \Pi_I^{\ j} \equiv 2 \pi^{jk}\,\delta_{IL}e^L_{\ k}\,,\\
& Q^{[MN]} = E^M_{\ i}\,\Sigma_K^{\ i}\,\delta^{KN}- E^N_{\ i}\,\Sigma_K^{\ i}\,\delta^{KM}\,, \qquad 
\Sigma_I^{\ j} \equiv 2 \sigma^{jk}\,\delta_{IL}E^L_{\ k}\,,
\end{align}
\end{subequations}
and $\mathcal{R}_0^\text{GR}$, $\mathcal{R}_i$, $\tilde{\mathcal{R}}_0^\text{GR}$ and $\tilde{\mathcal{R}}_i$ are given by~\eqref{eq:prec_contr}.

As those constraints have to be conserved in time, we have to check whether new constraints are necessary.
Imposing that $\dot{P}^{[MN]} \approx 0$ yields $Y^{[MN]} = 0$.
Similarly, imposing $\dot{Q}^{[MN]} \approx 0$ yields $X^{[MN]} = 0$.
As those conditions are redundant, it implies that 3 combinations of $P^{[MN]}$ and $Q^{[MN]}$ are first-class constraints (corresponding to simultaneous boosts of the two sets of vielbeins) and that we have to add 3 more constraints (e.g. $Y^{[MN]}$) to the secondary Hamiltonian.
As for the Poisson algebra of the eight first constraints, it is naturally still given by the matrix~(\ref{table:precursor}).
As the Poisson brackets of any of the eight first constraints with $P^{[MN]}$ or $Q^{[MN]}$ is vanishing, we can conclude that there are $4$ first-class constraints and $4$ second-class constraints among the eight first constraints.

As each vielbein bears 9 components, we have started with $2\times 2 \times 9 = 36$ phase space variables.
We have $3 + 4$ first-class constraints and $3+3+4$ second-class constraints, removing a total of 24 phase space \emph{dof}, so that we reach the same conclusion as in the metric formulation, namely that the precursor theory contains at most 6 physical \emph{dof}.

\subsection{The Minimal Theory of Bigravity}

The Hamiltonian of MTBG in the vielbein language reads
\begin{equation}
\begin{aligned}
& H = -\int\!\dd^3x\left\lbrace N\mathcal{R}_0 + N^i\mathcal{R}_i+M\tilde{\mathcal{R}}_0 + M^i\tilde{\mathcal{R}}_i
+ \alpha_{MN}P^{[MN]}+ \beta_{MN}Q^{[MN]}+ \lambda_{MN}Y^{[MN]}\right.\\
& \left. \hspace{3cm}
+ \left(\lambda+ \gamma^{ij}\mathcal{D}_{ij}\bar\lambda\right)C_0
- \left(\lambda - \phi^{ij}\mathfrak{D}_{ij}\bar\lambda\right)\tilde{C}_0 + \lambda^i\left(C_i- \beta\,\tilde{C}_i\right) \right\rbrace\,,
\end{aligned}
\end{equation}
where
\begin{subequations}
\begin{align}
& C_0 = 
\frac{m^2}{2}\,W_I^{\ J}\,\left(\gamma_{ik}\,E_J^{\ k}\,e^I_{\ j}+\gamma_{jk}\,E_J^{\ k}\,e^I_{\ i}-\gamma_{ij}Y_J^{\ I}\right)\pi^{ij}\,,\\
& C_i =
- \Mpl^2\,m^2\sqrt\gamma\,\mathcal{D}^j\left(W_I^{\ J}\,Y_J^{\ K}\,\delta_{KL}\,e^I_{\ i}\,e^L_{\ j}\right)\,,\\
& \tilde{C}_0 = 
\frac{m^2}{2\alpha^2}\,\tilde{W}_I^{\ J}\,\left(\phi_{ik}\,e_J^{\ k}\,E^I_{\ j}+\phi_{jk}\,e_J^{\ k}\,E^I_{\ i}-\phi_{ij}X_J^{\ I}\right)\sigma^{ij}\,,\\
& \tilde{C}_i =
- \Mpl^2\,m^2\sqrt\phi\,\mathfrak{D}^j\left(\tilde{W}_I^{\ J}\,X_J^{\ K}\,\delta_{KL}\,E^I_{\ i}\,E^L_{\ j}\right)\,,
\end{align}
\end{subequations}
and 
\begin{subequations}
\begin{align}
& W_I^{\ J} \equiv \frac{\sqrt\phi}{\sqrt\gamma}\left[c_1\,\delta^J_I + c_2\left(Y_K^{\ K}\,\delta_I^J-Y_I^{\ J}\right)\right]+c_3\,X_I^{\ J}\,,\\
& \tilde{W}_I^{\ J} \equiv \frac{\sqrt\gamma}{\sqrt\phi}\left[c_3\,\delta^J_I + c_2\left(X_K^{\ K}\,\delta_I^J-X_I^{\ J}\right)\right]+c_1\,Y_I^{\ J}\,.
\end{align}
\end{subequations}

The Hamiltonian equations of motion for $e^I_{\ j}$ and $E^I_{\ j}$ give the relations
\begin{subequations}
\begin{align}
&\frac{2}{\Mpl^2}\frac{\pi^{ij}}{\sqrt\gamma} = 
K^{ij} - K\,\gamma^{ij} - \frac{m^2M}{4N} \left(\lambda+ \gamma^{kl}\mathcal{D}_{kl}\bar\lambda\right)\,\Theta^{ij}\,,\\
&\frac{2}{\alpha^2\Mpl^2}\frac{\sigma^{ij}}{\sqrt\phi} = 
\Phi^{ij} - \Phi\,\phi^{ij} + \frac{m^2N}{4M} \left(\lambda-\phi^{kl}\mathfrak{D}_{kl}\bar\lambda\right)\,\tilde{\Theta}^{ij}\,,
\end{align}
\end{subequations}
where
\begin{equation}
\Theta^{ij} = W_I^{\ J}\,\delta^{IK}\left(e_K^{\ i}\,E_J^{\ j}+e_K^{\ j}\,E_J^{\ i}\right)\,,
\qquad\text{and}\qquad
\tilde{\Theta}^{ij} = \tilde{W}_I^{\ J}\,\delta^{IK}\left(E_K^{\ i}\,e_J^{\ j}+E_K^{\ j}\,e_J^{\ i}\right)\,.
\end{equation}
Performing a Legendre transformation, we obtain the Lagrangian density
\begin{equation}
\begin{aligned}
\mathcal{L} = & \:
\mathcal{L}_\text{pre} 
+ \lambda^i\left(C_i- \beta\,\tilde{C}_i\right)\\
& 
- \frac{m^2\Mpl^2}{2}\sqrt\gamma\left(\lambda + \gamma^{kl}\mathcal{D}_{kl}\bar\lambda\right) K^{ij}\gamma_{jk}E_J^{\ k}\,e^I_{\ j}\,W_I^{\ J}
+ \frac{m^2\Mpl^2}{2}\sqrt\phi\left(\lambda - \phi^{kl}\mathfrak{D}_{kl}\bar\lambda\right) \Phi^{ij}\phi_{jk}e_J^{\ k}\,E^I_{\ j}\,\tilde{W}_I^{\ J}\\
& 
+ \frac{\Mpl^2\,m^4}{32N}\sqrt\gamma\,\left(\lambda + \gamma^{pq}\mathcal{D}_{pq}\bar\lambda\right)^2\left(\gamma_{ik}\gamma_{jl}-\frac{1}{2}\,\gamma_{ij}\gamma_{kl}\right)\Theta^{ij}\,\Theta^{kl}\\
& 
+ \frac{\Mpl^2\,m^4}{32\alpha^2M}\sqrt\phi\,\left(\lambda - \phi^{pq}\mathfrak{D}_{pq}\bar\lambda\right)^2\left(\phi_{ik}\phi_{jl}-\frac{1}{2}\,\phi_{ij}\phi_{kl}\right)\tilde{\Theta}^{ij}\,\tilde{\Theta}^{kl}\,.
\end{aligned}
\end{equation}

\section{Correspondences between the notations of~\cite{bigrav_viable_cosmo} and those of the present work}\label{app:dic}

Table~\ref{table:dic} shows the correspondences between the notations in this work on MTBG and those used in~\cite{bigrav_viable_cosmo} on HRBG. In addition, \cite{bigrav_viable_cosmo} introduced a non-vanishing spatial curvature, $K$, while it is not the case in the present work.

\setlength{\tabcolsep}{9pt}
\begin{table}[h!]
\begin{center}
\begin{tabular}{|c||c|c|c|c|c|c|c|c|c|c|}
\hline
\cite{bigrav_viable_cosmo}'s notations & $N$ & $n$ & $a$ & $\alpha$ & $H$ & $H_f$ & $\xi$ & $\tilde{c}$ & $\kappa$ & $\alpha_n$\\ 
\hline
This work's notations & $N$ & $M$ & $a$ & $b$ & $H$ & $L$ & $\mathcal{X}$ & $r$ & $\alpha^2$ & $c_n$\\
\hline
\end{tabular}
\end{center}
\begin{center}
\begin{tabular}{|c||c|c|c|c|c|c|}
\hline
\cite{bigrav_viable_cosmo}'s notations & $m^2 \hat{\rho}_{m,g}$ & $m^2 \hat{\rho}_{m,f}$ & $\rho_g$ & $\rho_f$ & $-m^2 \xi J(\tilde{c}-1)$ & $\frac{m^2}{\xi^3 \tilde{c}}J(\tilde{c}-1)$ \\ 
\hline
This work's notations & $\frac{\rho_m}{\Mpl^2}$ & $\frac{\tilde{\rho}_m}{\Mpl^2}$ & $\rho$ & $\tilde{\rho}$ & $\frac{p_m+\rho_m}{\Mpl^2}$ & $\frac{\tilde{p}_m + \tilde{\rho}_m}{\Mpl^2}$ \\
\hline
\end{tabular}
\end{center}
\caption{Dictionary  between this work's notations and those in~\cite{bigrav_viable_cosmo}.}
\label{table:dic}
\end{table}

\section{Mini-superspace Hamiltonian treatment}\label{app:hamil_back}

Let us consider the background dynamics in the Hamiltonian formalism. From the Lagrangian in mini-superspace given in Eq.\ (\ref{eq:lag_mini}), we can extract the mini-superspace Hamiltonian (in the presence of two perfect fluids, one for each metric) as
\begin{eqnarray}
H & = & \frac{12\,\Mpl^{2}a^{4}\rho(J^{0})+6\,a{m}^{2}\Mpl^{2}\left(a^{3}c_{{4}}+3\,a^{2}b\,c_{{3}}+3\,ab^{2}c_{{2}}+b^{3}c_{{1}}\right)\Mpl^{2}-\pi_{a}^{2}}{12\,\Mpl^{2}a}\,N\nonumber \\
 & + & \frac{12\,\alpha^{2}\Mpl^{2}b^{4}\,\tilde{\rho}(\tilde{J}^{0})+6\,bm^{2}\Mpl^{2}\left(b^{3}c_{0}+3\,ab^{2}c_{{1}}+3\,a^{2}b\,c_{{2}}+a^{3}c_{{3}}\right)\alpha^{2}\Mpl^{2}-\pi_{b}^{2}}{12\,\alpha^{2}\Mpl^{2}b}\,M\nonumber \\
 & + & \frac{m^{2}\left(a^{2}c_{3}+2\,ab\,c_{2}+b^{2}c_{1}\right)\left(a\pi_{{\it af}}-\alpha^{2}b\,\pi_{{a}}\right)\lambda}{4\,\alpha^{2}\,a\,b}\nonumber \\
 & + & \left(\tilde{J}^{0}\,b^{3}+\pi_{\tilde{\phi}}\right)\lambda_{1}+\left(J^{0}a^{3}+\pi_{\phi}\right)\lambda_{2}+\lambda_{3}\pi_{J^{0}}+\lambda_{4}\pi_{\tilde{J}^{0}}\,.\label{eq:Ham}
\end{eqnarray}
We can then deduce the presence of the following constraints
\begin{eqnarray}
C_{N} & \equiv & \frac{12\,\Mpl^{2}a^{4}\rho(J^{0})+6\,a{m}^{2}\Mpl^{2}\left(a^{3}c_{{4}}+3\,a^{2}b\,c_{{3}}+3\,ab^{2}c_{{2}}+b^{3}c_{{1}}\right)\Mpl^{2}-\pi_{{a}}^{2}}{12\,\Mpl^{2}a}\,,\\
C_{M} & \equiv & \frac{12\,\alpha^{2}\Mpl^{2}b^{4}\,\tilde{\rho}(\tilde{J}^{0})+6\,bm^{2}\Mpl^{2}\left(b^{3}c_{0}+3\,ab^{2}c_{{1}}+3\,a^{2}b\,c_{{2}}+a^{3}c_{{3}}\right)\alpha^{2}\Mpl^{2}-\pi_{b}^{2}}{12\,\alpha^{2}\Mpl^{2}b}\,,\\
C_{\lambda} & \equiv & \frac{m^{2}\left(a^{2}c_{3}+2\,ab\,c_{2}+b^{2}c_{1}\right)\left(a\pi_{{\it b}}-\alpha^{2}b\,\pi_{a}\right)}{4\,\alpha^{2}\,a\,b}\,,\\
C_{1} & \equiv & \tilde{J}^{0}\,b^{3}+\pi_{\tilde{\phi}}\,,\\
C_{2} & \equiv & J^{0}\,a^{3}+\pi_{\phi}\,,\\
C_{3} & \equiv & \pi_{J^{0}}\,,\\
C_{4} & \equiv & \pi_{\tilde{J}^{0}}\,,
\end{eqnarray}
where $J^\mu=(J^0/N,\vec 0)$ and $\tilde{J}^\mu=(\tilde{J}^0/M,\vec 0)$ belong to the two perfect fluids endowed with mini-superspace Lagrangians $-a^3[N\rho(J^0) +J^0\dot\varphi]$ and $-b^3[M \tilde{\rho}(\tilde{J}^0) +\tilde{J}^0\dot{\tilde{\varphi}}] $ respectively.

We can immediately see that $C_{\lambda}\approx0$ leads to two different
possibilities for the constraint surface, which describe the two different
branches defined by setting
\begin{equation}
a^{2}c_{3}+2\,ab\,c_{2}+b^{2}c_{1}\approx0\,,\qquad{\rm or}\qquad a\pi_{{\it b}}-\alpha^{2}b\,\pi_{a}\approx0\,.
\end{equation}
Hence, we define the self-accelerating branch by setting $a^{2}c_{3}+2\,ab\,c_{2}+b^{2}c_{1}\approx0$
and the normal branch by the relation $a\pi_{{\it b}}-\alpha^{2}b\,\pi_{{a}}\approx0$.

The $C_{\lambda}\approx0$ constraint is exactly the same as the one
present in Hassan-Rosen bigravity (HRBG), but for HRBG, it
comes as a consequence of the time-derivatives of the Hamiltonian
constraints\footnote{To make comparison with the HRBG paper, one needs  to make the following constants redefinitions:
  $m_{g}=\Mpl/\sqrt{2}$, $m_{f}=\alpha \Mpl/\sqrt{2}$, $m_{{\rm HR}}^{4}=m^{2}\Mpl^{2}/4$, and
  $\beta_{n}=c_{4-n}.$}. 
In fact, the primary Hamiltonian for HRBG
 corresponds to the expression written in Eq.\
(\ref{eq:Ham}), on setting
$\lambda$ to vanish. However on taking the conditions
$\dot{C}_{N}^{{\rm HR}}\approx0$, and $\dot{C}_{M}^{{\rm
    HR}}\approx0$ (together with
$\dot{C}_i\approx0$ with
$i=1,\dots,4$) we obtain a secondary constraint, namely the condition
$C_{\lambda}\approx0$. In other words in HRBG,
$C_{\lambda}\approx0$ is obtained as a secondary constraint and
$\lambda$ is just introduced as a Lagrange multiplier (in mini superspace) used to implement such a new secondary
constraint. Therefore, on adding this secondary constraint to the HRBG
primary Hamiltonian, we find the two Hamiltonians, i.e.\ the one for
MTBG and the other of HRBG, are equivalent, and, as such, lead to the same background equations
of motion.

Let us finally discuss the value of $\lambda$ for both branches.
By taking the time derivative of the $C_{i}$ ($i=1,\dots,4$) constraints,
we obtain
\begin{eqnarray}
\dot{C}_{3} & \approx & 0 \Rightarrow \lambda_{2}=-N\frac{\partial\rho}{\partial J^{0}}\,,\\
\dot{C}_{4} & \approx & 0 \Rightarrow \lambda_{1}=-M\frac{\partial\tilde{\rho}}{\partial\tilde{J}^{0}}\,,\\
\dot{C}_{2} & \approx & 0 \Rightarrow \lambda_{3}=\frac{J^{0}\pi_{a}N}{2a^{2}\Mpl^{2}}+\frac{3m^{2}\lambda J^{0}}{4a^{2}}\,(c_{1}b^{2}+2c_{2}ab+c_{3}a^{2})\,,\\
\dot{C}_{1} & \approx & 0 \Rightarrow \lambda_{4}=\frac{\tilde{J}^{0}\pi_{b}M}{2b^{2}\alpha^{2}\Mpl^{2}}-\frac{3m^{2}\lambda\tilde{J}^{0}}{4b^{2}\alpha^{2}}\,(c_{1}b^{2}+2c_{2}ab+c_{3}a^{2})\,,
\end{eqnarray}
which set the values of the Lagrange multipliers $\lambda_{i}$ ($i=1,\dots,4$). On using these relations, we find that
\begin{equation}
\dot{C}_{N}\approx0 \Rightarrow F[\dots]\,\lambda\approx0\,,
\end{equation}
where $F$ is an expression which does not vanish, in general, in
any of the two branches. A new constraint of this kind is also found by considering
$\dot{C}_{M}\approx0$. Then, if we were to make these new constraints vanish 
by $not$ setting $\lambda$ to zero on the constraints surface,
we would too strongly constrain the dynamics of the background. Therefore
for both branches we will set
\begin{equation}
\lambda\approx0\,.
\end{equation}


\bibliography{MTBG_biblio}

\begin{thebibliography}{51}%
\makeatletter
\providecommand \@ifxundefined [1]{%
 \@ifx{#1\undefined}
}%
\providecommand \@ifnum [1]{%
 \ifnum #1\expandafter \@firstoftwo
 \else \expandafter \@secondoftwo
 \fi
}%
\providecommand \@ifx [1]{%
 \ifx #1\expandafter \@firstoftwo
 \else \expandafter \@secondoftwo
 \fi
}%
\providecommand \natexlab [1]{#1}%
\providecommand \enquote  [1]{``#1''}%
\providecommand \bibnamefont  [1]{#1}%
\providecommand \bibfnamefont [1]{#1}%
\providecommand \citenamefont [1]{#1}%
\providecommand \href@noop [0]{\@secondoftwo}%
\providecommand \href [0]{\begingroup \@sanitize@url \@href}%
\providecommand \@href[1]{\@@startlink{#1}\@@href}%
\providecommand \@@href[1]{\endgroup#1\@@endlink}%
\providecommand \@sanitize@url [0]{\catcode `\\12\catcode `\$12\catcode
  `\&12\catcode `\#12\catcode `\^12\catcode `\_12\catcode `\%12\relax}%
\providecommand \@@startlink[1]{}%
\providecommand \@@endlink[0]{}%
\providecommand \url  [0]{\begingroup\@sanitize@url \@url }%
\providecommand \@url [1]{\endgroup\@href {#1}{\urlprefix }}%
\providecommand \urlprefix  [0]{URL }%
\providecommand \Eprint [0]{\href }%
\providecommand \doibase [0]{https://doi.org/}%
\providecommand \selectlanguage [0]{\@gobble}%
\providecommand \bibinfo  [0]{\@secondoftwo}%
\providecommand \bibfield  [0]{\@secondoftwo}%
\providecommand \translation [1]{[#1]}%
\providecommand \BibitemOpen [0]{}%
\providecommand \bibitemStop [0]{}%
\providecommand \bibitemNoStop [0]{.\EOS\space}%
\providecommand \EOS [0]{\spacefactor3000\relax}%
\providecommand \BibitemShut  [1]{\csname bibitem#1\endcsname}%
\let\auto@bib@innerbib\@empty
\bibitem [{\citenamefont {Perlmutter}\ \emph {et~al.}(1999)\citenamefont
  {Perlmutter}, \citenamefont {Aldering}, \citenamefont {Goldhaber},
  \citenamefont {Knop}, \citenamefont {Nugent}, \citenamefont {Castro},
  \citenamefont {Deustua}, \citenamefont {Fabbro}, \citenamefont {Goobar},
  \citenamefont {Groom},\ and\ \citenamefont {et~al.}}]{DE_obs_1}%
  \BibitemOpen
  \bibfield  {author} {\bibinfo {author} {\bibfnamefont {S.}~\bibnamefont
  {Perlmutter}}, \bibinfo {author} {\bibfnamefont {G.}~\bibnamefont
  {Aldering}}, \bibinfo {author} {\bibfnamefont {G.}~\bibnamefont {Goldhaber}},
  \bibinfo {author} {\bibfnamefont {R.~A.}\ \bibnamefont {Knop}}, \bibinfo
  {author} {\bibfnamefont {P.}~\bibnamefont {Nugent}}, \bibinfo {author}
  {\bibfnamefont {P.~G.}\ \bibnamefont {Castro}}, \bibinfo {author}
  {\bibfnamefont {S.}~\bibnamefont {Deustua}}, \bibinfo {author} {\bibfnamefont
  {S.}~\bibnamefont {Fabbro}}, \bibinfo {author} {\bibfnamefont
  {A.}~\bibnamefont {Goobar}}, \bibinfo {author} {\bibfnamefont {D.~E.}\
  \bibnamefont {Groom}},\ and\ \bibinfo {author} {\bibnamefont {et~al.}},\
  }\bibfield  {title} {\bibinfo {title} {Measurements of omega and lambda from
  42 high-redshift supernovae},\ }\href {https://doi.org/10.1086/307221}
  {\bibfield  {journal} {\bibinfo  {journal} {The Astrophysical Journal}\
  }\textbf {\bibinfo {volume} {517}},\ \bibinfo {pages} {565} (\bibinfo {year}
  {1999})},\ \Eprint {https://arxiv.org/abs/astro-ph/9812133}
  {arXiv:astro-ph/9812133 [astro-ph]} \BibitemShut {NoStop}%
\bibitem [{\citenamefont {Riess}\ \emph {et~al.}(1998)\citenamefont {Riess},
  \citenamefont {Filippenko}, \citenamefont {Challis}, \citenamefont
  {Clocchiatti}, \citenamefont {Diercks}, \citenamefont {Garnavich},
  \citenamefont {Gilliland}, \citenamefont {Hogan}, \citenamefont {Jha},
  \citenamefont {Kirshner},\ and\ \citenamefont {et~al.}}]{DE_obs_2}%
  \BibitemOpen
  \bibfield  {author} {\bibinfo {author} {\bibfnamefont {A.~G.}\ \bibnamefont
  {Riess}}, \bibinfo {author} {\bibfnamefont {A.~V.}\ \bibnamefont
  {Filippenko}}, \bibinfo {author} {\bibfnamefont {P.}~\bibnamefont {Challis}},
  \bibinfo {author} {\bibfnamefont {A.}~\bibnamefont {Clocchiatti}}, \bibinfo
  {author} {\bibfnamefont {A.}~\bibnamefont {Diercks}}, \bibinfo {author}
  {\bibfnamefont {P.~M.}\ \bibnamefont {Garnavich}}, \bibinfo {author}
  {\bibfnamefont {R.~L.}\ \bibnamefont {Gilliland}}, \bibinfo {author}
  {\bibfnamefont {C.~J.}\ \bibnamefont {Hogan}}, \bibinfo {author}
  {\bibfnamefont {S.}~\bibnamefont {Jha}}, \bibinfo {author} {\bibfnamefont
  {R.~P.}\ \bibnamefont {Kirshner}},\ and\ \bibinfo {author} {\bibnamefont
  {et~al.}},\ }\bibfield  {title} {\bibinfo {title} {Observational evidence
  from supernovae for an accelerating universe and a cosmological constant},\
  }\href {https://doi.org/10.1086/300499} {\bibfield  {journal} {\bibinfo
  {journal} {The Astronomical Journal}\ }\textbf {\bibinfo {volume} {116}},\
  \bibinfo {pages} {1009} (\bibinfo {year} {1998})},\ \Eprint
  {https://arxiv.org/abs/astro-ph/9805201} {arXiv:astro-ph/9805201 [astro-ph]}
  \BibitemShut {NoStop}%
\bibitem [{\citenamefont {Abbott}\ \emph {et~al.}(2019)\citenamefont {Abbott},
  \citenamefont {Abbott}, \citenamefont {Abbott}, \citenamefont {Acernese},
  \citenamefont {Ackley}, \citenamefont {Adams}, \citenamefont {Adams},
  \citenamefont {Addesso}, \citenamefont {Adhikari}, \citenamefont {Adya},\
  and\ \citenamefont {et~al.}}]{GW170817_tests}%
  \BibitemOpen
  \bibfield  {author} {\bibinfo {author} {\bibfnamefont {B.}~\bibnamefont
  {Abbott}}, \bibinfo {author} {\bibfnamefont {R.}~\bibnamefont {Abbott}},
  \bibinfo {author} {\bibfnamefont {T.}~\bibnamefont {Abbott}}, \bibinfo
  {author} {\bibfnamefont {F.}~\bibnamefont {Acernese}}, \bibinfo {author}
  {\bibfnamefont {K.}~\bibnamefont {Ackley}}, \bibinfo {author} {\bibfnamefont
  {C.}~\bibnamefont {Adams}}, \bibinfo {author} {\bibfnamefont
  {T.}~\bibnamefont {Adams}}, \bibinfo {author} {\bibfnamefont
  {P.}~\bibnamefont {Addesso}}, \bibinfo {author} {\bibfnamefont
  {R.}~\bibnamefont {Adhikari}}, \bibinfo {author} {\bibfnamefont
  {V.}~\bibnamefont {Adya}},\ and\ \bibinfo {author} {\bibnamefont {et~al.}},\
  }\bibfield  {title} {\bibinfo {title} {Tests of general relativity with
  gw170817},\ }\bibfield  {journal} {\bibinfo  {journal} {Physical Review
  Letters}\ }\textbf {\bibinfo {volume} {123}},\ \href
  {https://doi.org/10.1103/physrevlett.123.011102}
  {10.1103/physrevlett.123.011102} (\bibinfo {year} {2019}),\ \Eprint
  {https://arxiv.org/abs/1811.00364} {arXiv:1811.00364 [gr-qc]} \BibitemShut
  {NoStop}%
\bibitem [{\citenamefont {Fierz}\ and\ \citenamefont
  {Pauli}(1939)}]{Fierz_Pauli}%
  \BibitemOpen
  \bibfield  {author} {\bibinfo {author} {\bibfnamefont {M.}~\bibnamefont
  {Fierz}}\ and\ \bibinfo {author} {\bibfnamefont {W.~E.}\ \bibnamefont
  {Pauli}},\ }\bibfield  {title} {\bibinfo {title} {On relativistic wave
  equations for particles of arbitrary spin in an electromagnetic field},\
  }\href {https://doi.org/10.1098/rspa.1939.0140} {\bibfield  {journal}
  {\bibinfo  {journal} {Proc. R. Soc. Lond. A}\ }\textbf {\bibinfo {volume}
  {173}},\ \bibinfo {pages} {211} (\bibinfo {year} {1939})}\BibitemShut
  {NoStop}%
\bibitem [{\citenamefont {de~Rham}\ \emph {et~al.}(2011)\citenamefont
  {de~Rham}, \citenamefont {Gabadadze},\ and\ \citenamefont {Tolley}}]{dRGT_1}%
  \BibitemOpen
  \bibfield  {author} {\bibinfo {author} {\bibfnamefont {C.}~\bibnamefont
  {de~Rham}}, \bibinfo {author} {\bibfnamefont {G.}~\bibnamefont {Gabadadze}},\
  and\ \bibinfo {author} {\bibfnamefont {A.~J.}\ \bibnamefont {Tolley}},\
  }\bibfield  {title} {\bibinfo {title} {Resummation of massive gravity},\
  }\bibfield  {journal} {\bibinfo  {journal} {Physical Review Letters}\
  }\textbf {\bibinfo {volume} {106}},\ \href
  {https://doi.org/10.1103/physrevlett.106.231101}
  {10.1103/physrevlett.106.231101} (\bibinfo {year} {2011}),\ \Eprint
  {https://arxiv.org/abs/1011.1232} {arXiv:1011.1232 [hep-th]} \BibitemShut
  {NoStop}%
\bibitem [{\citenamefont {de~Rham}(2014)}]{dRGT_2}%
  \BibitemOpen
  \bibfield  {author} {\bibinfo {author} {\bibfnamefont {C.}~\bibnamefont
  {de~Rham}},\ }\bibfield  {title} {\bibinfo {title} {Massive gravity},\
  }\bibfield  {journal} {\bibinfo  {journal} {Living Reviews in Relativity}\
  }\textbf {\bibinfo {volume} {17}},\ \href
  {https://doi.org/10.12942/lrr-2014-7} {10.12942/lrr-2014-7} (\bibinfo {year}
  {2014})\BibitemShut {NoStop}%
\bibitem [{\citenamefont {De~Felice}\ \emph {et~al.}(2012)\citenamefont
  {De~Felice}, \citenamefont {Gumrukcuoglu},\ and\ \citenamefont
  {Mukohyama}}]{dRGT_no_FLRW}%
  \BibitemOpen
  \bibfield  {author} {\bibinfo {author} {\bibfnamefont {A.}~\bibnamefont
  {De~Felice}}, \bibinfo {author} {\bibfnamefont {A.~E.}\ \bibnamefont
  {Gumrukcuoglu}},\ and\ \bibinfo {author} {\bibfnamefont {S.}~\bibnamefont
  {Mukohyama}},\ }\bibfield  {title} {\bibinfo {title} {Massive gravity:
  Nonlinear instability of a homogeneous and isotropic universe},\ }\bibfield
  {journal} {\bibinfo  {journal} {Physical Review Letters}\ }\textbf {\bibinfo
  {volume} {109}},\ \href {https://doi.org/10.1103/physrevlett.109.171101}
  {10.1103/physrevlett.109.171101} (\bibinfo {year} {2012}),\ \Eprint
  {https://arxiv.org/abs/1206.2080} {arXiv:1206.2080 [hep-th]} \BibitemShut
  {NoStop}%
\bibitem [{\citenamefont {Hassan}\ and\ \citenamefont
  {Rosen}(2012)}]{Hassan_Rosen}%
  \BibitemOpen
  \bibfield  {author} {\bibinfo {author} {\bibfnamefont {S.~F.}\ \bibnamefont
  {Hassan}}\ and\ \bibinfo {author} {\bibfnamefont {R.~A.}\ \bibnamefont
  {Rosen}},\ }\bibfield  {title} {\bibinfo {title} {Bimetric gravity from
  ghost-free massive gravity},\ }\bibfield  {journal} {\bibinfo  {journal}
  {Journal of High Energy Physics}\ }\textbf {\bibinfo {volume} {2012}},\ \href
  {https://doi.org/10.1007/jhep02(2012)126} {10.1007/jhep02(2012)126} (\bibinfo
  {year} {2012}),\ \Eprint {https://arxiv.org/abs/1109.3515} {arXiv:1109.3515
  [hep-th]} \BibitemShut {NoStop}%
\bibitem [{\citenamefont {Comelli}\ \emph {et~al.}(2012)\citenamefont
  {Comelli}, \citenamefont {Crisostomi},\ and\ \citenamefont
  {Pilo}}]{bigrav_inst_1}%
  \BibitemOpen
  \bibfield  {author} {\bibinfo {author} {\bibfnamefont {D.}~\bibnamefont
  {Comelli}}, \bibinfo {author} {\bibfnamefont {M.}~\bibnamefont
  {Crisostomi}},\ and\ \bibinfo {author} {\bibfnamefont {L.}~\bibnamefont
  {Pilo}},\ }\bibfield  {title} {\bibinfo {title} {Perturbations in massive
  gravity cosmology},\ }\bibfield  {journal} {\bibinfo  {journal} {Journal of
  High Energy Physics}\ }\textbf {\bibinfo {volume} {2012}},\ \href
  {https://doi.org/10.1007/jhep06(2012)085} {10.1007/jhep06(2012)085} (\bibinfo
  {year} {2012}),\ \Eprint {https://arxiv.org/abs/1202.1986} {arXiv:1202.1986
  [hep-th]} \BibitemShut {NoStop}%
\bibitem [{\citenamefont {Koennig}\ and\ \citenamefont
  {Amendola}(2014)}]{bigrav_inst_2}%
  \BibitemOpen
  \bibfield  {author} {\bibinfo {author} {\bibfnamefont {F.}~\bibnamefont
  {Koennig}}\ and\ \bibinfo {author} {\bibfnamefont {L.}~\bibnamefont
  {Amendola}},\ }\bibfield  {title} {\bibinfo {title} {Instability in a minimal
  bimetric gravity model},\ }\bibfield  {journal} {\bibinfo  {journal}
  {Physical Review D}\ }\textbf {\bibinfo {volume} {90}},\ \href
  {https://doi.org/10.1103/physrevd.90.044030} {10.1103/physrevd.90.044030}
  (\bibinfo {year} {2014}),\ \Eprint {https://arxiv.org/abs/1402.1988}
  {arXiv:1402.1988 [astro-ph.CO]} \BibitemShut {NoStop}%
\bibitem [{\citenamefont {Akrami}\ \emph {et~al.}(2015)\citenamefont {Akrami},
  \citenamefont {Hassan}, \citenamefont {K{\"o}nnig}, \citenamefont
  {Schmidt-May},\ and\ \citenamefont {Solomon}}]{bigrav_Planck_hierarchy}%
  \BibitemOpen
  \bibfield  {author} {\bibinfo {author} {\bibfnamefont {Y.}~\bibnamefont
  {Akrami}}, \bibinfo {author} {\bibfnamefont {S.}~\bibnamefont {Hassan}},
  \bibinfo {author} {\bibfnamefont {F.}~\bibnamefont {K{\"o}nnig}}, \bibinfo
  {author} {\bibfnamefont {A.}~\bibnamefont {Schmidt-May}},\ and\ \bibinfo
  {author} {\bibfnamefont {A.~R.}\ \bibnamefont {Solomon}},\ }\bibfield
  {title} {\bibinfo {title} {Bimetric gravity is cosmologically viable},\
  }\href {https://doi.org/10.1016/j.physletb.2015.06.062} {\bibfield  {journal}
  {\bibinfo  {journal} {Physics Letters B}\ }\textbf {\bibinfo {volume}
  {748}},\ \bibinfo {pages} {37} (\bibinfo {year} {2015})},\ \Eprint
  {https://arxiv.org/abs/1503.07521} {arXiv:1503.07521 [gr-qc]} \BibitemShut
  {NoStop}%
\bibitem [{\citenamefont {De~Felice}\ \emph
  {et~al.}(2018{\natexlab{a}})\citenamefont {De~Felice}, \citenamefont
  {Mukohyama},\ and\ \citenamefont {Uzan}}]{Cham_bigrav_OP}%
  \BibitemOpen
  \bibfield  {author} {\bibinfo {author} {\bibfnamefont {A.}~\bibnamefont
  {De~Felice}}, \bibinfo {author} {\bibfnamefont {S.}~\bibnamefont
  {Mukohyama}},\ and\ \bibinfo {author} {\bibfnamefont {J.-P.}\ \bibnamefont
  {Uzan}},\ }\bibfield  {title} {\bibinfo {title} {Extending applicability of
  bimetric theory: chameleon bigravity},\ }\bibfield  {journal} {\bibinfo
  {journal} {General Relativity and Gravitation}\ }\textbf {\bibinfo {volume}
  {50}},\ \href {https://doi.org/10.1007/s10714-018-2342-z}
  {10.1007/s10714-018-2342-z} (\bibinfo {year} {2018}{\natexlab{a}}),\ \Eprint
  {https://arxiv.org/abs/1702.04490} {arXiv:1702.04490 [hep-th]} \BibitemShut
  {NoStop}%
\bibitem [{\citenamefont {De~Felice}\ \emph
  {et~al.}(2018{\natexlab{b}})\citenamefont {De~Felice}, \citenamefont
  {Mukohyama}, \citenamefont {Oliosi},\ and\ \citenamefont
  {Watanabe}}]{Cham_bigrav_cosmo}%
  \BibitemOpen
  \bibfield  {author} {\bibinfo {author} {\bibfnamefont {A.}~\bibnamefont
  {De~Felice}}, \bibinfo {author} {\bibfnamefont {S.}~\bibnamefont
  {Mukohyama}}, \bibinfo {author} {\bibfnamefont {M.}~\bibnamefont {Oliosi}},\
  and\ \bibinfo {author} {\bibfnamefont {Y.}~\bibnamefont {Watanabe}},\
  }\bibfield  {title} {\bibinfo {title} {Stable cosmology in chameleon
  bigravity},\ }\bibfield  {journal} {\bibinfo  {journal} {Physical Review D}\
  }\textbf {\bibinfo {volume} {97}},\ \href
  {https://doi.org/10.1103/physrevd.97.024050} {10.1103/physrevd.97.024050}
  (\bibinfo {year} {2018}{\natexlab{b}}),\ \Eprint
  {https://arxiv.org/abs/1711.04655} {arXiv:1711.04655 [hep-th]} \BibitemShut
  {NoStop}%
\bibitem [{\citenamefont {de~Rham}\ \emph
  {et~al.}(2014{\natexlab{a}})\citenamefont {de~Rham}, \citenamefont
  {Keltner},\ and\ \citenamefont {Tolley}}]{GMG_OP}%
  \BibitemOpen
  \bibfield  {author} {\bibinfo {author} {\bibfnamefont {C.}~\bibnamefont
  {de~Rham}}, \bibinfo {author} {\bibfnamefont {L.}~\bibnamefont {Keltner}},\
  and\ \bibinfo {author} {\bibfnamefont {A.~J.}\ \bibnamefont {Tolley}},\
  }\bibfield  {title} {\bibinfo {title} {Generalized galileon duality},\
  }\bibfield  {journal} {\bibinfo  {journal} {Physical Review D}\ }\textbf
  {\bibinfo {volume} {90}},\ \href {https://doi.org/10.1103/physrevd.90.024050}
  {10.1103/physrevd.90.024050} (\bibinfo {year} {2014}{\natexlab{a}}),\ \Eprint
  {https://arxiv.org/abs/1403.3690} {arXiv:1403.3690 [hep-th]} \BibitemShut
  {NoStop}%
\bibitem [{\citenamefont {de~Rham}\ \emph
  {et~al.}(2014{\natexlab{b}})\citenamefont {de~Rham}, \citenamefont
  {Fasiello},\ and\ \citenamefont {Tolley}}]{GMG_cosmo_1}%
  \BibitemOpen
  \bibfield  {author} {\bibinfo {author} {\bibfnamefont {C.}~\bibnamefont
  {de~Rham}}, \bibinfo {author} {\bibfnamefont {M.}~\bibnamefont {Fasiello}},\
  and\ \bibinfo {author} {\bibfnamefont {A.~J.}\ \bibnamefont {Tolley}},\
  }\bibfield  {title} {\bibinfo {title} {Stable flrw solutions in generalized
  massive gravity},\ }\href {https://doi.org/10.1142/s0218271814430068}
  {\bibfield  {journal} {\bibinfo  {journal} {International Journal of Modern
  Physics D}\ }\textbf {\bibinfo {volume} {23}},\ \bibinfo {pages} {1443006}
  (\bibinfo {year} {2014}{\natexlab{b}})},\ \Eprint
  {https://arxiv.org/abs/1410.0960} {arXiv:1410.0960 [hep-th]} \BibitemShut
  {NoStop}%
\bibitem [{\citenamefont {Kenna-Allison}\ \emph {et~al.}(2020)\citenamefont
  {Kenna-Allison}, \citenamefont {Gumrukcuoglu},\ and\ \citenamefont
  {Koyama}}]{GMG_cosmo_2}%
  \BibitemOpen
  \bibfield  {author} {\bibinfo {author} {\bibfnamefont {M.}~\bibnamefont
  {Kenna-Allison}}, \bibinfo {author} {\bibfnamefont {A.~E.}\ \bibnamefont
  {Gumrukcuoglu}},\ and\ \bibinfo {author} {\bibfnamefont {K.}~\bibnamefont
  {Koyama}},\ }\bibfield  {title} {\bibinfo {title} {Stable cosmology in
  generalized massive gravity},\ }\bibfield  {journal} {\bibinfo  {journal}
  {Physical Review D}\ }\textbf {\bibinfo {volume} {101}},\ \href
  {https://doi.org/10.1103/physrevd.101.084014} {10.1103/physrevd.101.084014}
  (\bibinfo {year} {2020}),\ \Eprint {https://arxiv.org/abs/1912.08560}
  {arXiv:1912.08560 [hep-th]} \BibitemShut {NoStop}%
\bibitem [{\citenamefont {G\"{u}mr\"{u}k\c{c}\"{u}o\v{g}lu}\ \emph
  {et~al.}(2020)\citenamefont {G\"{u}mr\"{u}k\c{c}\"{u}o\v{g}lu}, \citenamefont
  {Kimura},\ and\ \citenamefont {Koyama}}]{Non_minimal_coupling}%
  \BibitemOpen
  \bibfield  {author} {\bibinfo {author} {\bibfnamefont {A.~E.}\ \bibnamefont
  {G\"{u}mr\"{u}k\c{c}\"{u}o\v{g}lu}}, \bibinfo {author} {\bibfnamefont
  {R.}~\bibnamefont {Kimura}},\ and\ \bibinfo {author} {\bibfnamefont
  {K.}~\bibnamefont {Koyama}},\ }\href
  {https://doi.org/10.1103/physrevd.101.124021} {\bibinfo {title} {Massive
  gravity with nonminimal coupling}} (\bibinfo {year} {2020}),\ \Eprint
  {https://arxiv.org/abs/2003.11831} {arXiv:2003.11831 [gr-qc]} \BibitemShut
  {NoStop}%
\bibitem [{\citenamefont {Will}(2014)}]{Will_pulsars}%
  \BibitemOpen
  \bibfield  {author} {\bibinfo {author} {\bibfnamefont {C.~M.}\ \bibnamefont
  {Will}},\ }\bibfield  {title} {\bibinfo {title} {The confrontation between
  general relativity and experiment},\ }\bibfield  {journal} {\bibinfo
  {journal} {Living Reviews in Relativity}\ }\textbf {\bibinfo {volume} {17}},\
  \href {https://doi.org/10.12942/lrr-2014-4} {10.12942/lrr-2014-4} (\bibinfo
  {year} {2014}),\ \Eprint {https://arxiv.org/abs/1403.7377} {arXiv:1403.7377
  [gr-qc]} \BibitemShut {NoStop}%
\bibitem [{\citenamefont {Abbott}\ \emph {et~al.}(2017)\citenamefont {Abbott},
  \citenamefont {Abbott}, \citenamefont {Abbott}, \citenamefont {Acernese},
  \citenamefont {Ackley}, \citenamefont {Adams}, \citenamefont {Adams},
  \citenamefont {Addesso}, \citenamefont {Adhikari}, \citenamefont {Adya},\
  and\ \citenamefont {et~al.}}]{GW170814_pol}%
  \BibitemOpen
  \bibfield  {author} {\bibinfo {author} {\bibfnamefont {B.}~\bibnamefont
  {Abbott}}, \bibinfo {author} {\bibfnamefont {R.}~\bibnamefont {Abbott}},
  \bibinfo {author} {\bibfnamefont {T.}~\bibnamefont {Abbott}}, \bibinfo
  {author} {\bibfnamefont {F.}~\bibnamefont {Acernese}}, \bibinfo {author}
  {\bibfnamefont {K.}~\bibnamefont {Ackley}}, \bibinfo {author} {\bibfnamefont
  {C.}~\bibnamefont {Adams}}, \bibinfo {author} {\bibfnamefont
  {T.}~\bibnamefont {Adams}}, \bibinfo {author} {\bibfnamefont
  {P.}~\bibnamefont {Addesso}}, \bibinfo {author} {\bibfnamefont
  {R.}~\bibnamefont {Adhikari}}, \bibinfo {author} {\bibfnamefont
  {V.}~\bibnamefont {Adya}},\ and\ \bibinfo {author} {\bibnamefont {et~al.}},\
  }\bibfield  {title} {\bibinfo {title} {Gw170814: A three-detector observation
  of gravitational waves from a binary black hole coalescence},\ }\bibfield
  {journal} {\bibinfo  {journal} {Physical Review Letters}\ }\textbf {\bibinfo
  {volume} {119}},\ \href {https://doi.org/10.1103/physrevlett.119.141101}
  {10.1103/physrevlett.119.141101} (\bibinfo {year} {2017}),\ \Eprint
  {https://arxiv.org/abs/1709.09660} {arXiv:1709.09660 [gr-qc]} \BibitemShut
  {NoStop}%
\bibitem [{\citenamefont {Baumgartner}\ \emph {et~al.}(2020)\citenamefont
  {Baumgartner}, \citenamefont {Bernardini}, \citenamefont {Canivete~Cuissa},
  \citenamefont {de~Laroussilhe}, \citenamefont {Mitchell}, \citenamefont
  {Neuenschwander}, \citenamefont {Saha}, \citenamefont {Schaeffer},
  \citenamefont {Soyuer},\ and\ \citenamefont {Zwick}}]{pol_LISA}%
  \BibitemOpen
  \bibfield  {author} {\bibinfo {author} {\bibfnamefont {S.}~\bibnamefont
  {Baumgartner}}, \bibinfo {author} {\bibfnamefont {M.}~\bibnamefont
  {Bernardini}}, \bibinfo {author} {\bibfnamefont {J.~R.}\ \bibnamefont
  {Canivete~Cuissa}}, \bibinfo {author} {\bibfnamefont {H.}~\bibnamefont
  {de~Laroussilhe}}, \bibinfo {author} {\bibfnamefont {A.~M.~W.}\ \bibnamefont
  {Mitchell}}, \bibinfo {author} {\bibfnamefont {B.~A.}\ \bibnamefont
  {Neuenschwander}}, \bibinfo {author} {\bibfnamefont {P.}~\bibnamefont
  {Saha}}, \bibinfo {author} {\bibfnamefont {T.}~\bibnamefont {Schaeffer}},
  \bibinfo {author} {\bibfnamefont {D.}~\bibnamefont {Soyuer}},\ and\ \bibinfo
  {author} {\bibfnamefont {L.}~\bibnamefont {Zwick}},\ }\bibfield  {title}
  {\bibinfo {title} {Towards a polarization prediction for lisa via intensity
  interferometry},\ }\href {https://doi.org/10.1093/mnras/staa2638} {\bibfield
  {journal} {\bibinfo  {journal} {Monthly Notices of the Royal Astronomical
  Society}\ }\textbf {\bibinfo {volume} {498}},\ \bibinfo {pages} {4577}
  (\bibinfo {year} {2020})},\ \Eprint {https://arxiv.org/abs/2008.11538}
  {arXiv:2008.11538 [astro-ph]} \BibitemShut {NoStop}%
\bibitem [{\citenamefont {De~Felice}\ and\ \citenamefont
  {Mukohyama}(2016{\natexlab{a}})}]{MTMG_OP}%
  \BibitemOpen
  \bibfield  {author} {\bibinfo {author} {\bibfnamefont {A.}~\bibnamefont
  {De~Felice}}\ and\ \bibinfo {author} {\bibfnamefont {S.}~\bibnamefont
  {Mukohyama}},\ }\bibfield  {title} {\bibinfo {title} {Minimal theory of
  massive gravity},\ }\href@noop {} {\bibfield  {journal} {\bibinfo  {journal}
  {Physics Letters B}\ }\textbf {\bibinfo {volume} {752}},\ \bibinfo {pages}
  {302} (\bibinfo {year} {2016}{\natexlab{a}})},\ \Eprint
  {https://arxiv.org/abs/1506.01594} {arXiv:1506.01594 [gr-qc]} \BibitemShut
  {NoStop}%
\bibitem [{\citenamefont {De~Felice}\ and\ \citenamefont
  {Mukohyama}(2016{\natexlab{b}})}]{MTMG_pheno}%
  \BibitemOpen
  \bibfield  {author} {\bibinfo {author} {\bibfnamefont {A.}~\bibnamefont
  {De~Felice}}\ and\ \bibinfo {author} {\bibfnamefont {S.}~\bibnamefont
  {Mukohyama}},\ }\bibfield  {title} {\bibinfo {title} {Phenomenology in
  minimal theory of massive gravity},\ }\href@noop {} {\bibfield  {journal}
  {\bibinfo  {journal} {JCAP}\ }\textbf {\bibinfo {volume} {04}},\ \bibinfo
  {pages} {028}},\ \Eprint {https://arxiv.org/abs/1512.04008} {arXiv:1512.04008
  [gr-qc]} \BibitemShut {NoStop}%
\bibitem [{\citenamefont {De~Felice}\ \emph
  {et~al.}(2018{\natexlab{c}})\citenamefont {De~Felice}, \citenamefont
  {Larrouturou}, \citenamefont {Mukohyama},\ and\ \citenamefont
  {Oliosi}}]{MTMG_BH}%
  \BibitemOpen
  \bibfield  {author} {\bibinfo {author} {\bibfnamefont {A.}~\bibnamefont
  {De~Felice}}, \bibinfo {author} {\bibfnamefont {F.}~\bibnamefont
  {Larrouturou}}, \bibinfo {author} {\bibfnamefont {S.}~\bibnamefont
  {Mukohyama}},\ and\ \bibinfo {author} {\bibfnamefont {M.}~\bibnamefont
  {Oliosi}},\ }\bibfield  {title} {\bibinfo {title} {Black holes and stars in
  the minimal theory of massive gravity},\ }\href@noop {} {\bibfield  {journal}
  {\bibinfo  {journal} {Phys. Rev. D}\ }\textbf {\bibinfo {volume} {98}},\
  \bibinfo {pages} {104031} (\bibinfo {year} {2018}{\natexlab{c}})},\ \Eprint
  {https://arxiv.org/abs/1808.01403} {arXiv:1808.01403 [gr-qc]} \BibitemShut
  {NoStop}%
\bibitem [{\citenamefont {De~Felice}\ and\ \citenamefont
  {Mukohyama}(2017)}]{MTMG_RSD}%
  \BibitemOpen
  \bibfield  {author} {\bibinfo {author} {\bibfnamefont {A.}~\bibnamefont
  {De~Felice}}\ and\ \bibinfo {author} {\bibfnamefont {S.}~\bibnamefont
  {Mukohyama}},\ }\bibfield  {title} {\bibinfo {title} {{Graviton mass might
  reduce tension between early and late time cosmological data}},\ }\href
  {https://doi.org/10.1103/PhysRevLett.118.091104} {\bibfield  {journal}
  {\bibinfo  {journal} {Phys. Rev. Lett.}\ }\textbf {\bibinfo {volume} {118}},\
  \bibinfo {pages} {091104} (\bibinfo {year} {2017})},\ \Eprint
  {https://arxiv.org/abs/1607.03368} {arXiv:1607.03368 [astro-ph.CO]}
  \BibitemShut {NoStop}%
\bibitem [{\citenamefont {Bolis}\ \emph {et~al.}(2018)\citenamefont {Bolis},
  \citenamefont {De~Felice},\ and\ \citenamefont {Mukohyama}}]{MTMG_ISW}%
  \BibitemOpen
  \bibfield  {author} {\bibinfo {author} {\bibfnamefont {N.}~\bibnamefont
  {Bolis}}, \bibinfo {author} {\bibfnamefont {A.}~\bibnamefont {De~Felice}},\
  and\ \bibinfo {author} {\bibfnamefont {S.}~\bibnamefont {Mukohyama}},\
  }\bibfield  {title} {\bibinfo {title} {Integrated sachs-wolfe-galaxy
  cross-correlation bounds on the two branches of the minimal theory of massive
  gravity},\ }\bibfield  {journal} {\bibinfo  {journal} {Phys. Rev. D}\
  }\textbf {\bibinfo {volume} {98}},\ \href
  {https://doi.org/10.1103/physrevd.98.024010} {10.1103/physrevd.98.024010}
  (\bibinfo {year} {2018}),\ \Eprint {https://arxiv.org/abs/1804.01790}
  {arXiv:1804.01790 [astro-ph]} \BibitemShut {NoStop}%
\bibitem [{\citenamefont {Hagala}\ \emph {et~al.}(2020)\citenamefont {Hagala},
  \citenamefont {Felice}, \citenamefont {Mota},\ and\ \citenamefont
  {Mukohyama}}]{MTMG_Nbody}%
  \BibitemOpen
  \bibfield  {author} {\bibinfo {author} {\bibfnamefont {R.}~\bibnamefont
  {Hagala}}, \bibinfo {author} {\bibfnamefont {A.~D.}\ \bibnamefont {Felice}},
  \bibinfo {author} {\bibfnamefont {D.~F.}\ \bibnamefont {Mota}},\ and\
  \bibinfo {author} {\bibfnamefont {S.}~\bibnamefont {Mukohyama}},\ }\href@noop
  {} {\bibinfo {title} {Nonlinear dynamics of the minimal theory of massive
  gravity}} (\bibinfo {year} {2020}),\ \Eprint
  {https://arxiv.org/abs/2011.14697} {arXiv:2011.14697 [astro-ph.CO]}
  \BibitemShut {NoStop}%
\bibitem [{\citenamefont {De~Felice}\ \emph
  {et~al.}(2017{\natexlab{a}})\citenamefont {De~Felice}, \citenamefont
  {Mukohyama},\ and\ \citenamefont {Oliosi}}]{MQD_OP}%
  \BibitemOpen
  \bibfield  {author} {\bibinfo {author} {\bibfnamefont {A.}~\bibnamefont
  {De~Felice}}, \bibinfo {author} {\bibfnamefont {S.}~\bibnamefont
  {Mukohyama}},\ and\ \bibinfo {author} {\bibfnamefont {M.}~\bibnamefont
  {Oliosi}},\ }\bibfield  {title} {\bibinfo {title} {Minimal theory of
  quasidilaton massive gravity},\ }\bibfield  {journal} {\bibinfo  {journal}
  {Phys. Rev. D}\ }\textbf {\bibinfo {volume} {96}},\ \href
  {https://doi.org/10.1103/physrevd.96.024032} {10.1103/physrevd.96.024032}
  (\bibinfo {year} {2017}{\natexlab{a}}),\ \Eprint
  {https://arxiv.org/abs/1701.01581} {arXiv:1701.01581 [hep-th]} \BibitemShut
  {NoStop}%
\bibitem [{\citenamefont {De~Felice}\ \emph
  {et~al.}(2017{\natexlab{b}})\citenamefont {De~Felice}, \citenamefont
  {Mukohyama},\ and\ \citenamefont {Oliosi}}]{MQD_Horndeski}%
  \BibitemOpen
  \bibfield  {author} {\bibinfo {author} {\bibfnamefont {A.}~\bibnamefont
  {De~Felice}}, \bibinfo {author} {\bibfnamefont {S.}~\bibnamefont
  {Mukohyama}},\ and\ \bibinfo {author} {\bibfnamefont {M.}~\bibnamefont
  {Oliosi}},\ }\bibfield  {title} {\bibinfo {title} {Horndeski extension of the
  minimal theory of quasidilaton massive gravity},\ }\bibfield  {journal}
  {\bibinfo  {journal} {Phys. Rev. D}\ }\textbf {\bibinfo {volume} {96}},\
  \href {https://doi.org/10.1103/physrevd.96.104036}
  {10.1103/physrevd.96.104036} (\bibinfo {year} {2017}{\natexlab{b}}),\ \Eprint
  {https://arxiv.org/abs/1709.03108} {arXiv:1709.03108 [hep-th]} \BibitemShut
  {NoStop}%
\bibitem [{\citenamefont {De~Felice}\ \emph {et~al.}(2019)\citenamefont
  {De~Felice}, \citenamefont {Mukohyama},\ and\ \citenamefont
  {Oliosi}}]{MQD_pheno}%
  \BibitemOpen
  \bibfield  {author} {\bibinfo {author} {\bibfnamefont {A.}~\bibnamefont
  {De~Felice}}, \bibinfo {author} {\bibfnamefont {S.}~\bibnamefont
  {Mukohyama}},\ and\ \bibinfo {author} {\bibfnamefont {M.}~\bibnamefont
  {Oliosi}},\ }\bibfield  {title} {\bibinfo {title} {Phenomenology of minimal
  theory of quasidilaton massive gravity},\ }\bibfield  {journal} {\bibinfo
  {journal} {Physical Review D}\ }\textbf {\bibinfo {volume} {99}},\ \href
  {https://doi.org/10.1103/physrevd.99.044055} {10.1103/physrevd.99.044055}
  (\bibinfo {year} {2019}),\ \Eprint {https://arxiv.org/abs/1806.00602}
  {arXiv:1806.00602 [hep-th]} \BibitemShut {NoStop}%
\bibitem [{\citenamefont {Lin}\ and\ \citenamefont
  {Mukohyama}(2017)}]{MMTG_OP}%
  \BibitemOpen
  \bibfield  {author} {\bibinfo {author} {\bibfnamefont {C.}~\bibnamefont
  {Lin}}\ and\ \bibinfo {author} {\bibfnamefont {S.}~\bibnamefont
  {Mukohyama}},\ }\bibfield  {title} {\bibinfo {title} {A class of minimally
  modified gravity theories},\ }\href
  {https://doi.org/10.1088/1475-7516/2017/10/033} {\bibfield  {journal}
  {\bibinfo  {journal} {Journal of Cosmology and Astroparticle Physics}\
  }\textbf {\bibinfo {volume} {2017}}\bibfield  {number} {\bibinfo  {number} {
  (10)},\ \bibinfo {pages} {033}},\ }\Eprint {https://arxiv.org/abs/1708.03757}
  {arXiv:1708.03757 [gr-qc]} \BibitemShut {NoStop}%
\bibitem [{\citenamefont {Carballo-Rubio}\ \emph {et~al.}(2018)\citenamefont
  {Carballo-Rubio}, \citenamefont {Filippo},\ and\ \citenamefont
  {Liberati}}]{MMTG_Carballo_Rubio}%
  \BibitemOpen
  \bibfield  {author} {\bibinfo {author} {\bibfnamefont {R.}~\bibnamefont
  {Carballo-Rubio}}, \bibinfo {author} {\bibfnamefont {F.~D.}\ \bibnamefont
  {Filippo}},\ and\ \bibinfo {author} {\bibfnamefont {S.}~\bibnamefont
  {Liberati}},\ }\bibfield  {title} {\bibinfo {title} {Minimally modified
  theories of gravity: a playground for testing the uniqueness of general
  relativity},\ }\href {https://doi.org/10.1088/1475-7516/2018/06/026}
  {\bibfield  {journal} {\bibinfo  {journal} {Journal of Cosmology and
  Astroparticle Physics}\ }\textbf {\bibinfo {volume} {2018}}\bibfield
  {number} {\bibinfo  {number} { (06)},\ \bibinfo {pages} {026}},\ }\Eprint
  {https://arxiv.org/abs/1802.02537} {arXiv:1802.02537 [gr-qc]} \BibitemShut
  {NoStop}%
\bibitem [{\citenamefont {Lin}(2019)}]{MMTG_Lin}%
  \BibitemOpen
  \bibfield  {author} {\bibinfo {author} {\bibfnamefont {C.}~\bibnamefont
  {Lin}},\ }\bibfield  {title} {\bibinfo {title} {The self-consistent matter
  coupling of a class of minimally modified gravity theories},\ }\href
  {https://doi.org/10.1088/1475-7516/2019/05/037} {\bibfield  {journal}
  {\bibinfo  {journal} {Journal of Cosmology and Astroparticle Physics}\
  }\textbf {\bibinfo {volume} {2019}}\bibfield  {number} {\bibinfo  {number} {
  (05)},\ \bibinfo {pages} {037}},\ }\Eprint {https://arxiv.org/abs/1811.02467}
  {arXiv:1811.02467 [gr-qc]} \BibitemShut {NoStop}%
\bibitem [{\citenamefont {Mukohyama}\ and\ \citenamefont
  {Noui}(2019)}]{MMTG_Hamiltonian}%
  \BibitemOpen
  \bibfield  {author} {\bibinfo {author} {\bibfnamefont {S.}~\bibnamefont
  {Mukohyama}}\ and\ \bibinfo {author} {\bibfnamefont {K.}~\bibnamefont
  {Noui}},\ }\bibfield  {title} {\bibinfo {title} {Minimally modified gravity:
  a hamiltonian construction},\ }\href
  {https://doi.org/10.1088/1475-7516/2019/07/049} {\bibfield  {journal}
  {\bibinfo  {journal} {Journal of Cosmology and Astroparticle Physics}\
  }\textbf {\bibinfo {volume} {2019}}\bibfield  {number} {\bibinfo  {number} {
  (07)},\ \bibinfo {pages} {049}},\ }\Eprint {https://arxiv.org/abs/1905.02000}
  {arXiv:1905.02000 [gr-qc]} \BibitemShut {NoStop}%
\bibitem [{\citenamefont {Aoki}\ \emph {et~al.}(2019)\citenamefont {Aoki},
  \citenamefont {Felice}, \citenamefont {Lin}, \citenamefont {Mukohyama},\ and\
  \citenamefont {Oliosi}}]{MMTG_pheno}%
  \BibitemOpen
  \bibfield  {author} {\bibinfo {author} {\bibfnamefont {K.}~\bibnamefont
  {Aoki}}, \bibinfo {author} {\bibfnamefont {A.~D.}\ \bibnamefont {Felice}},
  \bibinfo {author} {\bibfnamefont {C.}~\bibnamefont {Lin}}, \bibinfo {author}
  {\bibfnamefont {S.}~\bibnamefont {Mukohyama}},\ and\ \bibinfo {author}
  {\bibfnamefont {M.}~\bibnamefont {Oliosi}},\ }\bibfield  {title} {\bibinfo
  {title} {Phenomenology in type-\uppercase{I} minimally modified gravity},\
  }\href {https://doi.org/10.1088/1475-7516/2019/01/017} {\bibfield  {journal}
  {\bibinfo  {journal} {Journal of Cosmology and Astroparticle Physics}\
  }\textbf {\bibinfo {volume} {2019}}\bibfield  {number} {\bibinfo  {number} {
  (01)},\ \bibinfo {pages} {017}},\ }\Eprint {https://arxiv.org/abs/1810.01047}
  {arXiv:1810.01047 [gr-qc]} \BibitemShut {NoStop}%
\bibitem [{\citenamefont {Aoki}\ \emph {et~al.}(2020)\citenamefont {Aoki},
  \citenamefont {De~Felice}, \citenamefont {Mukohyama}, \citenamefont {Noui},
  \citenamefont {Oliosi},\ and\ \citenamefont {C.~Pookkillath}}]{MMTG_Planck}%
  \BibitemOpen
  \bibfield  {author} {\bibinfo {author} {\bibfnamefont {K.}~\bibnamefont
  {Aoki}}, \bibinfo {author} {\bibfnamefont {A.}~\bibnamefont {De~Felice}},
  \bibinfo {author} {\bibfnamefont {S.}~\bibnamefont {Mukohyama}}, \bibinfo
  {author} {\bibfnamefont {K.}~\bibnamefont {Noui}}, \bibinfo {author}
  {\bibfnamefont {M.}~\bibnamefont {Oliosi}},\ and\ \bibinfo {author}
  {\bibfnamefont {M.}~\bibnamefont {C.~Pookkillath}},\ }\bibfield  {title}
  {\bibinfo {title} {Minimally modified gravity fitting planck data better than
  $\lambda$-cdm},\ }\bibfield  {journal} {\bibinfo  {journal} {The European
  Physical Journal C}\ }\textbf {\bibinfo {volume} {80}},\ \href
  {https://doi.org/10.1140/epjc/s10052-020-8291-1}
  {10.1140/epjc/s10052-020-8291-1} (\bibinfo {year} {2020}),\ \Eprint
  {https://arxiv.org/abs/2005.13972} {arXiv:2005.13972 [astro-ph.CO]}
  \BibitemShut {NoStop}%
\bibitem [{\citenamefont {De~Felice}\ \emph {et~al.}(2020)\citenamefont
  {De~Felice}, \citenamefont {Doll}, \citenamefont {Larrouturou},\ and\
  \citenamefont {Mukohyama}}]{MMG2_BH}%
  \BibitemOpen
  \bibfield  {author} {\bibinfo {author} {\bibfnamefont {A.}~\bibnamefont
  {De~Felice}}, \bibinfo {author} {\bibfnamefont {A.}~\bibnamefont {Doll}},
  \bibinfo {author} {\bibfnamefont {F.}~\bibnamefont {Larrouturou}},\ and\
  \bibinfo {author} {\bibfnamefont {S.}~\bibnamefont {Mukohyama}},\ }\href@noop
  {} {\bibinfo {title} {{Black holes in a type-\uppercase{II} minimally
  modified gravity}}} (\bibinfo {year} {2020}),\ \Eprint
  {https://arxiv.org/abs/2010.13067} {arXiv:2010.13067 [gr-qc]} \BibitemShut
  {NoStop}%
\bibitem [{\citenamefont {Yao}\ \emph {et~al.}(2020)\citenamefont {Yao},
  \citenamefont {Oliosi}, \citenamefont {Gao},\ and\ \citenamefont
  {Mukohyama}}]{Yao:2020tur}%
  \BibitemOpen
  \bibfield  {author} {\bibinfo {author} {\bibfnamefont {Z.-B.}\ \bibnamefont
  {Yao}}, \bibinfo {author} {\bibfnamefont {M.}~\bibnamefont {Oliosi}},
  \bibinfo {author} {\bibfnamefont {X.}~\bibnamefont {Gao}},\ and\ \bibinfo
  {author} {\bibfnamefont {S.}~\bibnamefont {Mukohyama}},\ }\href@noop {}
  {\bibinfo {title} {{Minimally modified gravity with an auxiliary constraint:
  a Hamiltonian construction}}} (\bibinfo {year} {2020}),\ \Eprint
  {https://arxiv.org/abs/2011.00805} {arXiv:2011.00805 [gr-qc]} \BibitemShut
  {NoStop}%
\bibitem [{\citenamefont {Felice}\ and\ \citenamefont
  {Mukohyama}(2020)}]{MMG2_weakening}%
  \BibitemOpen
  \bibfield  {author} {\bibinfo {author} {\bibfnamefont {A.~D.}\ \bibnamefont
  {Felice}}\ and\ \bibinfo {author} {\bibfnamefont {S.}~\bibnamefont
  {Mukohyama}},\ }\href@noop {} {\bibinfo {title} {Weakening gravity for dark
  matter in a type-\uppercase{II} minimally modified gravity}} (\bibinfo {year}
  {2020}),\ \Eprint {https://arxiv.org/abs/2011.04188} {arXiv:2011.04188
  [astro-ph.CO]} \BibitemShut {NoStop}%
\bibitem [{\citenamefont {De~Felice}\ \emph
  {et~al.}(2014{\natexlab{a}})\citenamefont {De~Felice}, \citenamefont
  {G\"{u}mr\"{u}k\c{c}\"{u}o\v{g}lu}, \citenamefont {Mukohyama}, \citenamefont
  {Tanahashi},\ and\ \citenamefont {Tanaka}}]{bigrav_viable_cosmo}%
  \BibitemOpen
  \bibfield  {author} {\bibinfo {author} {\bibfnamefont {A.}~\bibnamefont
  {De~Felice}}, \bibinfo {author} {\bibfnamefont {A.~E.}\ \bibnamefont
  {G\"{u}mr\"{u}k\c{c}\"{u}o\v{g}lu}}, \bibinfo {author} {\bibfnamefont
  {S.}~\bibnamefont {Mukohyama}}, \bibinfo {author} {\bibfnamefont
  {N.}~\bibnamefont {Tanahashi}},\ and\ \bibinfo {author} {\bibfnamefont
  {T.}~\bibnamefont {Tanaka}},\ }\bibfield  {title} {\bibinfo {title} {Viable
  cosmology in bimetric theory},\ }\href
  {https://doi.org/10.1088/1475-7516/2014/06/037} {\bibfield  {journal}
  {\bibinfo  {journal} {Journal of Cosmology and Astroparticle Physics}\
  }\textbf {\bibinfo {volume} {2014}}\bibfield  {number} {\bibinfo  {number} {
  (06)},\ \bibinfo {pages} {037}},\ }\Eprint {https://arxiv.org/abs/1404.0008}
  {arXiv:1404.0008 [hep-th]} \BibitemShut {NoStop}%
\bibitem [{\citenamefont {Dirac}(2001)}]{Dirac}%
  \BibitemOpen
  \bibfield  {author} {\bibinfo {author} {\bibfnamefont {P.}~\bibnamefont
  {Dirac}},\ }\href {https://books.google.fr/books?id=GVwzb1rZW9kC} {\emph
  {\bibinfo {title} {Lectures on Quantum Mechanics}}},\ Belfer Graduate School
  of Science, monograph series\ (\bibinfo  {publisher} {Dover Publications},\
  \bibinfo {year} {2001})\BibitemShut {NoStop}%
\bibitem [{\citenamefont {Yamashita}\ \emph {et~al.}(2014)\citenamefont
  {Yamashita}, \citenamefont {De~Felice},\ and\ \citenamefont
  {Tanaka}}]{Yamashita}%
  \BibitemOpen
  \bibfield  {author} {\bibinfo {author} {\bibfnamefont {Y.}~\bibnamefont
  {Yamashita}}, \bibinfo {author} {\bibfnamefont {A.}~\bibnamefont
  {De~Felice}},\ and\ \bibinfo {author} {\bibfnamefont {T.}~\bibnamefont
  {Tanaka}},\ }\bibfield  {title} {\bibinfo {title} {Appearance of
  boulware-deser ghost in bigravity with doubly coupled matter},\ }\href
  {https://doi.org/10.1142/s0218271814430032} {\bibfield  {journal} {\bibinfo
  {journal} {International Journal of Modern Physics D}\ }\textbf {\bibinfo
  {volume} {23}},\ \bibinfo {pages} {1443003} (\bibinfo {year} {2014})},\
  \Eprint {https://arxiv.org/abs/1408.0487} {arXiv:1408.0487 [hep-th]}
  \BibitemShut {NoStop}%
\bibitem [{\citenamefont {Schutz}\ and\ \citenamefont
  {Sorkin}(1977)}]{Schutz77}%
  \BibitemOpen
  \bibfield  {author} {\bibinfo {author} {\bibfnamefont {B.~F.}\ \bibnamefont
  {Schutz}}\ and\ \bibinfo {author} {\bibfnamefont {R.}~\bibnamefont
  {Sorkin}},\ }\bibfield  {title} {\bibinfo {title} {{Variational aspects of
  relativistic field theories, with application to perfect fluids}},\ }\href
  {https://doi.org/10.1016/0003-4916(77)90200-7} {\bibfield  {journal}
  {\bibinfo  {journal} {Annals Phys.}\ }\textbf {\bibinfo {volume} {107}},\
  \bibinfo {pages} {1} (\bibinfo {year} {1977})}\BibitemShut {NoStop}%
\bibitem [{\citenamefont {Brown}(1993)}]{Brown93}%
  \BibitemOpen
  \bibfield  {author} {\bibinfo {author} {\bibfnamefont {J.~D.}\ \bibnamefont
  {Brown}},\ }\bibfield  {title} {\bibinfo {title} {Action functionals for
  relativistic perfect fluids},\ }\href
  {https://doi.org/10.1088/0264-9381/10/8/017} {\bibfield  {journal} {\bibinfo
  {journal} {Classical and Quantum Gravity}\ }\textbf {\bibinfo {volume}
  {10}},\ \bibinfo {pages} {1579} (\bibinfo {year} {1993})},\ \Eprint
  {https://arxiv.org/abs/gr-qc/9304026} {arXiv:gr-qc/9304026} \BibitemShut
  {NoStop}%
\bibitem [{\citenamefont {Moore}\ and\ \citenamefont
  {Nelson}(2001)}]{Grav_Cherenkov}%
  \BibitemOpen
  \bibfield  {author} {\bibinfo {author} {\bibfnamefont {G.~D.}\ \bibnamefont
  {Moore}}\ and\ \bibinfo {author} {\bibfnamefont {A.~E.}\ \bibnamefont
  {Nelson}},\ }\bibfield  {title} {\bibinfo {title} {{Lower bound on the
  propagation speed of gravity from gravitational Cherenkov radiation}},\
  }\href {https://doi.org/10.1088/1126-6708/2001/09/023} {\bibfield  {journal}
  {\bibinfo  {journal} {JHEP}\ }\textbf {\bibinfo {volume} {09}},\ \bibinfo
  {pages} {023}},\ \Eprint {https://arxiv.org/abs/hep-ph/0106220}
  {arXiv:hep-ph/0106220} \BibitemShut {NoStop}%
\bibitem [{\citenamefont {Kimura}\ \emph {et~al.}(2016)\citenamefont {Kimura},
  \citenamefont {Tanaka}, \citenamefont {Yamamoto},\ and\ \citenamefont
  {Yamashita}}]{bigravity_Cherenkov}%
  \BibitemOpen
  \bibfield  {author} {\bibinfo {author} {\bibfnamefont {R.}~\bibnamefont
  {Kimura}}, \bibinfo {author} {\bibfnamefont {T.}~\bibnamefont {Tanaka}},
  \bibinfo {author} {\bibfnamefont {K.}~\bibnamefont {Yamamoto}},\ and\
  \bibinfo {author} {\bibfnamefont {Y.}~\bibnamefont {Yamashita}},\ }\bibfield
  {title} {\bibinfo {title} {{Constraint on ghost-free bigravity from
  gravitational Cherenkov radiation}},\ }\href
  {https://doi.org/10.1103/PhysRevD.94.064059} {\bibfield  {journal} {\bibinfo
  {journal} {Phys. Rev. D}\ }\textbf {\bibinfo {volume} {94}},\ \bibinfo
  {pages} {064059} (\bibinfo {year} {2016})},\ \Eprint
  {https://arxiv.org/abs/1605.03405} {arXiv:1605.03405 [gr-qc]} \BibitemShut
  {NoStop}%
\bibitem [{\citenamefont {De~Felice}\ \emph
  {et~al.}(2014{\natexlab{b}})\citenamefont {De~Felice}, \citenamefont
  {Nakamura},\ and\ \citenamefont {Tanaka}}]{DeFelice:2013nba}%
  \BibitemOpen
  \bibfield  {author} {\bibinfo {author} {\bibfnamefont {A.}~\bibnamefont
  {De~Felice}}, \bibinfo {author} {\bibfnamefont {T.}~\bibnamefont
  {Nakamura}},\ and\ \bibinfo {author} {\bibfnamefont {T.}~\bibnamefont
  {Tanaka}},\ }\bibfield  {title} {\bibinfo {title} {{Possible existence of
  viable models of bi-gravity with detectable graviton oscillations by
  gravitational wave detectors}},\ }\href {https://doi.org/10.1093/ptep/ptu024}
  {\bibfield  {journal} {\bibinfo  {journal} {PTEP}\ }\textbf {\bibinfo
  {volume} {2014}},\ \bibinfo {pages} {043E01} (\bibinfo {year}
  {2014}{\natexlab{b}})},\ \Eprint {https://arxiv.org/abs/1304.3920}
  {arXiv:1304.3920 [gr-qc]} \BibitemShut {NoStop}%
\bibitem [{\citenamefont {Aoki}\ and\ \citenamefont
  {Mukohyama}(2016)}]{Aoki:2016zgp}%
  \BibitemOpen
  \bibfield  {author} {\bibinfo {author} {\bibfnamefont {K.}~\bibnamefont
  {Aoki}}\ and\ \bibinfo {author} {\bibfnamefont {S.}~\bibnamefont
  {Mukohyama}},\ }\bibfield  {title} {\bibinfo {title} {{Massive gravitons as
  dark matter and gravitational waves}},\ }\href
  {https://doi.org/10.1103/PhysRevD.94.024001} {\bibfield  {journal} {\bibinfo
  {journal} {Phys. Rev. D}\ }\textbf {\bibinfo {volume} {94}},\ \bibinfo
  {pages} {024001} (\bibinfo {year} {2016})},\ \Eprint
  {https://arxiv.org/abs/1604.06704} {arXiv:1604.06704 [hep-th]} \BibitemShut
  {NoStop}%
\bibitem [{\citenamefont {Fujita}\ \emph {et~al.}(2019)\citenamefont {Fujita},
  \citenamefont {Kuroyanagi}, \citenamefont {Mizuno},\ and\ \citenamefont
  {Mukohyama}}]{Fujita:2018ehq}%
  \BibitemOpen
  \bibfield  {author} {\bibinfo {author} {\bibfnamefont {T.}~\bibnamefont
  {Fujita}}, \bibinfo {author} {\bibfnamefont {S.}~\bibnamefont {Kuroyanagi}},
  \bibinfo {author} {\bibfnamefont {S.}~\bibnamefont {Mizuno}},\ and\ \bibinfo
  {author} {\bibfnamefont {S.}~\bibnamefont {Mukohyama}},\ }\bibfield  {title}
  {\bibinfo {title} {{Blue-tilted Primordial Gravitational Waves from Massive
  Gravity}},\ }\href {https://doi.org/10.1016/j.physletb.2018.12.025}
  {\bibfield  {journal} {\bibinfo  {journal} {Phys. Lett. B}\ }\textbf
  {\bibinfo {volume} {789}},\ \bibinfo {pages} {215} (\bibinfo {year}
  {2019})},\ \Eprint {https://arxiv.org/abs/1808.02381} {arXiv:1808.02381
  [gr-qc]} \BibitemShut {NoStop}%
\bibitem [{\citenamefont {de~Rham}\ \emph {et~al.}(2015)\citenamefont
  {de~Rham}, \citenamefont {Heisenberg},\ and\ \citenamefont
  {Ribeiro}}]{deRham:2014naa}%
  \BibitemOpen
  \bibfield  {author} {\bibinfo {author} {\bibfnamefont {C.}~\bibnamefont
  {de~Rham}}, \bibinfo {author} {\bibfnamefont {L.}~\bibnamefont
  {Heisenberg}},\ and\ \bibinfo {author} {\bibfnamefont {R.~H.}\ \bibnamefont
  {Ribeiro}},\ }\bibfield  {title} {\bibinfo {title} {{On couplings to matter
  in massive (bi-)gravity}},\ }\href
  {https://doi.org/10.1088/0264-9381/32/3/035022} {\bibfield  {journal}
  {\bibinfo  {journal} {Class. Quant. Grav.}\ }\textbf {\bibinfo {volume}
  {32}},\ \bibinfo {pages} {035022} (\bibinfo {year} {2015})},\ \Eprint
  {https://arxiv.org/abs/1408.1678} {arXiv:1408.1678 [hep-th]} \BibitemShut
  {NoStop}%
\bibitem [{\citenamefont {Gumrukcuoglu}\ \emph {et~al.}(2015)\citenamefont
  {Gumrukcuoglu}, \citenamefont {Heisenberg}, \citenamefont {Mukohyama},\ and\
  \citenamefont {Tanahashi}}]{Gumrukcuoglu:2015nua}%
  \BibitemOpen
  \bibfield  {author} {\bibinfo {author} {\bibfnamefont {A.~E.}\ \bibnamefont
  {Gumrukcuoglu}}, \bibinfo {author} {\bibfnamefont {L.}~\bibnamefont
  {Heisenberg}}, \bibinfo {author} {\bibfnamefont {S.}~\bibnamefont
  {Mukohyama}},\ and\ \bibinfo {author} {\bibfnamefont {N.}~\bibnamefont
  {Tanahashi}},\ }\bibfield  {title} {\bibinfo {title} {{Cosmology in bimetric
  theory with an effective composite coupling to matter}},\ }\href
  {https://doi.org/10.1088/1475-7516/2015/04/008} {\bibfield  {journal}
  {\bibinfo  {journal} {JCAP}\ }\textbf {\bibinfo {volume} {04}},\ \bibinfo
  {pages} {008}},\ \Eprint {https://arxiv.org/abs/1501.02790} {arXiv:1501.02790
  [hep-th]} \BibitemShut {NoStop}%
\bibitem [{\citenamefont {De~Felice}\ \emph {et~al.}(2016)\citenamefont
  {De~Felice}, \citenamefont {G\"umr\"uk\c{c}\"uo\u{g}lu}, \citenamefont
  {Heisenberg},\ and\ \citenamefont {Mukohyama}}]{DeFelice:2015yha}%
  \BibitemOpen
  \bibfield  {author} {\bibinfo {author} {\bibfnamefont {A.}~\bibnamefont
  {De~Felice}}, \bibinfo {author} {\bibfnamefont {A.~E.}\ \bibnamefont
  {G\"umr\"uk\c{c}\"uo\u{g}lu}}, \bibinfo {author} {\bibfnamefont
  {L.}~\bibnamefont {Heisenberg}},\ and\ \bibinfo {author} {\bibfnamefont
  {S.}~\bibnamefont {Mukohyama}},\ }\bibfield  {title} {\bibinfo {title}
  {{Matter coupling in partially constrained vielbein formulation of massive
  gravity}},\ }\href {https://doi.org/10.1088/1475-7516/2016/01/003} {\bibfield
   {journal} {\bibinfo  {journal} {JCAP}\ }\textbf {\bibinfo {volume} {01}},\
  \bibinfo {pages} {003}},\ \Eprint {https://arxiv.org/abs/1509.05978}
  {arXiv:1509.05978 [hep-th]} \BibitemShut {NoStop}%
\end{thebibliography}%

\end{document}